\documentclass[sigconf, nonacm]{acmart}

\AtBeginDocument{%
  \providecommand\BibTeX{{%
    \normalfont B\kern-0.5em{\scshape i\kern-0.25em b}\kern-0.8em\TeX}}}

\setcopyright{acmcopyright}
\copyrightyear{2018}
\acmYear{2018}
\acmDOI{XXXXXXX.XXXXXXX}

\acmConference[Preprint]{xxx}{In submission}{2026}
\acmBooktitle{Submission} 
\acmPrice{15.00}
\acmISBN{978-1-4503-XXXX-X/18/06}

\usepackage{hyperref}
\usepackage{enumitem, array}
\usepackage{xspace}
\usepackage{xcolor}
\usepackage{wrapfig}
\usepackage{csquotes}
\usepackage{comment}
\usepackage[capitalise, noabbrev]{cleveref}
\usepackage{caption}
\usepackage{subcaption}
\usepackage{multirow}
\usepackage{colortbl}
\usepackage{listings}
\usepackage{color-edits}
\addauthor{cm}{cyan}
\addauthor{shamsi}{teal}
\addauthor{sid}{orange}
\addauthor{pranav}{violet}
\addauthor{rock}{blue}
\addauthor{sw}{olive}
\addauthor{ken}{purple}
\addauthor{todo}{red}
\usepackage[most]{tcolorbox}
\usepackage{xcolor}

\newtcolorbox{taskbox}{
    colback=black!3,
    colframe=black!20,
    boxrule=0.4pt,
    arc=0mm,
    left=4mm,
    right=4mm,
    top=3mm,
    bottom=3mm,
    before skip=8pt,
    after skip=8pt,
    breakable
}

\newcommand{\paragraphBold}[1]{\paragraph{\emph{\textbf{#1}}}\xspace}

\newcolumntype{L}[1]{>{\raggedright\let\newline\\\arraybackslash\hspace{0pt}}m{#1}}
\newcolumntype{C}[1]{>{\centering\let\newline\\\arraybackslash\hspace{0pt}}m{#1}}
\newcolumntype{R}[1]{>{\raggedleft\let\newline\\\arraybackslash\hspace{0pt}}m{#1}}

\definecolor{cprompt}{HTML}{3c4043}
\definecolor{cinput}{HTML}{3182bd}
\definecolor{coutput}{HTML}{137333}

\definecolor{caddback}{rgb}{0.90, 0.98, 0.96}
\definecolor{cadd}{rgb}{0, 0.47, 0.34}
\definecolor{cdelback}{rgb}{1, 0.94, 0.92}
\definecolor{cdel}{rgb}{0.83, 0.32, 0.16}

\definecolor{cquote}{HTML}{3c4043}
\newcommand{\quoteinline}[1]{{\color{cquote}\emph{``#1'’}\xspace}}
\newcommand{\task}[1]{%
  {\texttt{#1}}\xspace}

\begin{document}

\title{Impact of Expertise and Trust on AI-assisted Work}

\author{Qianou Ma}
\authornote{Work done at Microsoft internship.}
\affiliation{%
  \institution{Carnegie Mellon University}
  \orcid{0009-0002-8634-130X}
  \city{Pittsburgh}
  \state{PA}
  \country{USA}
  \postcode{15213}
}
\email{qianouma@cmu.edu}

\author{Pranav Khadpe}
\affiliation{%
  \institution{Microsoft}
  \city{Redmond}
  \state{WA}
  \country{USA}
}
\email{prkhadpe@microsoft.com}

\author{Rock Yuren Pang}
\affiliation{%
  \institution{Microsoft}
  \city{Redmond}
  \state{WA}
  \country{USA}
}
\email{rockpang@microsoft.com}

\author{Kenneth Koedinger}
\affiliation{%
  \institution{Carnegie Mellon University}
  \city{Pittsburgh}
  \state{PA}
  \country{USA}
  \postcode{15213}
}
\email{koedinger@cmu.edu}

\author{Tongshuang Wu}
\affiliation{%
  \institution{Carnegie Mellon University}
  \city{Pittsburgh}
  \state{PA}
  \country{USA}
  \postcode{15213}
}
\email{sherryw@cs.cmu.edu}

\author{Siddharth Suri
}
\affiliation{%
  \institution{Microsoft Research}
  \city{Redmond}
  \state{WA}
  \country{USA}
}
\email{suri@microsoft.com}

\author{Shamsi Iqbal}
\affiliation{%
  \institution{Microsoft}
  \city{Redmond}
  \state{WA}
  \country{USA}
}
\email{shamsi@microsoft.com}

\renewcommand{\shortauthors}{Ma, et al.}

\begin{abstract}
Artificial intelligence (AI) agents promise to extend what users can accomplish beyond their existing expertise, but they also raise the risk of producing plausible-looking work that users may struggle to evaluate. We ask how task-relevant expertise and trust in AI shape agent-assisted work. To isolate these factors, we conduct a think-aloud study in which 20 professionals with varying data-science expertise use the same agentic tool to complete the same data-science hiring assessment, used at a major U.S. company. We find that surface productivity can come at the expense of understanding: participants who delegated more completed more of the task, but were also more likely to produce invalid solutions. Experts more often resisted this tradeoff, sacrificing completion to decompose the task and scrutinize agent outputs. Our findings call for evaluating agentic assistance not only by how much it accomplishes, but by whether users remain able to evaluate the work they produce.

\end{abstract}




\begin{teaserfigure}
    \centering
    \includegraphics[trim={0cm 0cm 0cm 0cm}, clip,width=\linewidth]{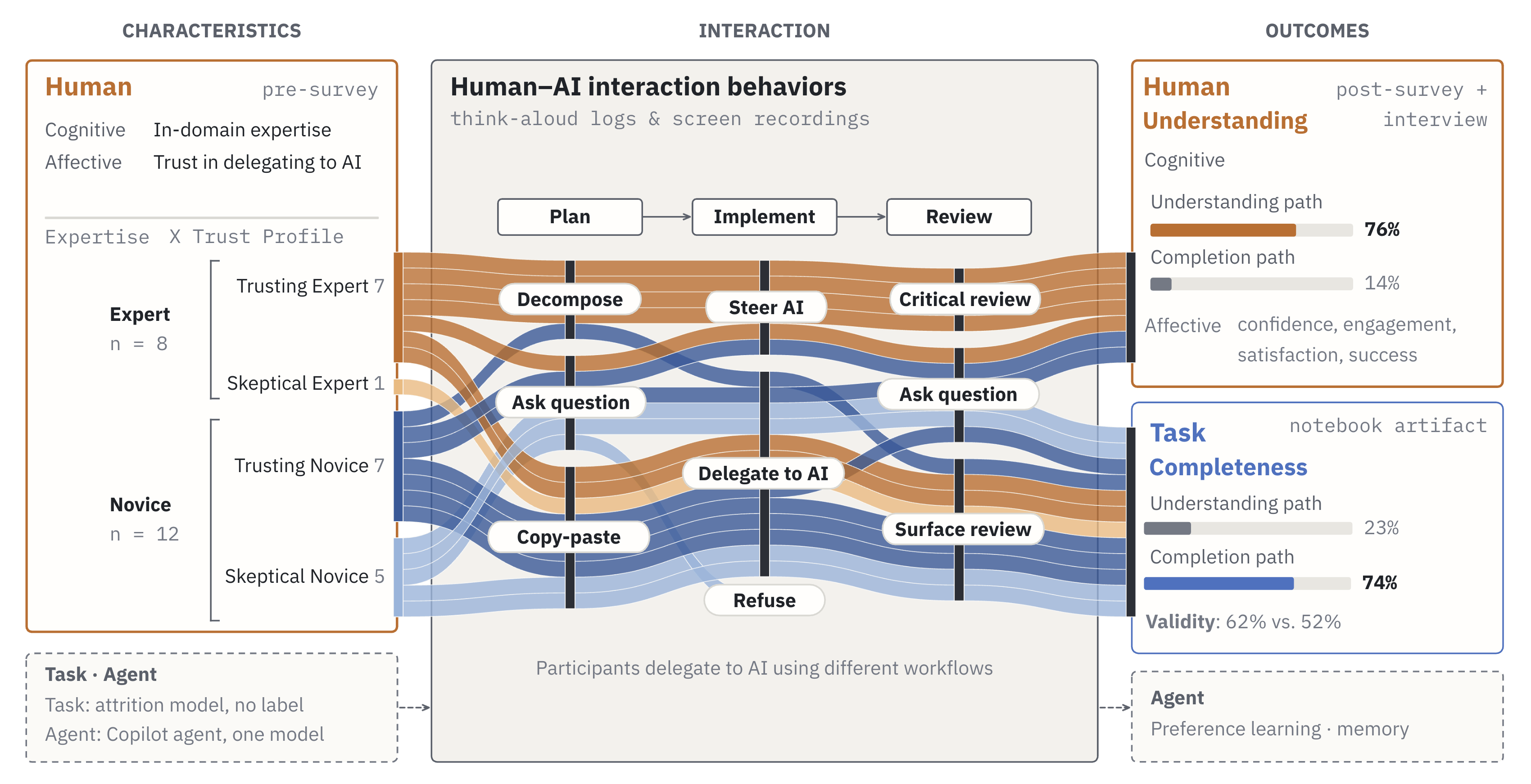} 
    \vspace{-15pt}
  \caption{
  Overview of our study framework and main findings. 
  We study how expertise and trust shape AI-assisted work by following 20 professionals from human \textit{characteristics}, through interaction  \textit{behaviors}, to human and task \textit{outcomes}.
  Experts more often decompose, steer, and critically review, while novices more often copy-paste, delegate to AI, and surface-review. These workflows reveal a central tradeoff: workflows that preserve human understanding tend to sacrifice task completion. 
  }
  \Description{...}
  \label{fig:ui}
\end{teaserfigure}

\maketitle

\section{Introduction}
\label{sec:intro}
General-purpose AI agents built on large language models (LLMs) are moving beyond answering questions to carrying out multi-step work, for example, writing code, analyzing data, and creating presentations. As these agents take on more of a task, the human's role shifts from \emph{doing} the work to \emph{steering} it: framing the problem, deciding what to delegate to an agent, and verifying what comes back \cite{Ma2026-uz}. This shift delivers on a central promise of generative AI in that it enables people to take on tasks beyond their existing expertise \cite{Daepp2026}. For example, a product manager can "vibe code" an interactive prototype by describing its desired behavior for an agent to help implement rather than relying entirely on an engineer to implement it \cite{Li2026-vibecoding}. More generally, an analysis of workplace ChatGPT use found that 16.8\% of workplace ChatGPT use involves people taking on work outside their own occupation~\cite{OpenAI2026-taskcrossover}.

Yet widespread concern about ``workslop'', i.e., AI-generated work that looks complete but is low-quality~\cite{liebscher2026workslop}, raises questions about how effectively people can steer and evaluate AI-assisted work, particularly outside their domain of expertise. AI-assisted deliverables routinely contain fabricated citations~\cite{walters2023fabrication}, implausible analyses~\cite{rawat2026plausible}, or factual errors~\cite{latif2025hallucinations}. This suggests that even when an artifact appears complete, workers may not understand it well enough to verify its quality, modify it, or carry it forward into subsequent work. The workslop concern is especially acute when working with agents because they are explicitly designed both to expand the range of tasks people can take on and to carry those tasks out with less oversight~\cite{franklin2026ai}. Contemporary agent interfaces increasingly make it easy for users to waive approvals, suppress interruptions, and cede oversight to the agent \cite{anthropic2026claudecodeautomode, mitchell2026ai}. Subsequently, users are encouraged to delegate work further beyond their expertise while also being nudged toward placing greater trust in an agent's ability to execute that work correctly~\cite{yang2026ai}.

To design agents that extend what workers can accomplish with AI while reducing the risk of low-quality outputs, we need to understand what shapes workers' use of these systems, particularly when they work on unfamiliar tasks. We therefore ask: \textbf{how do a worker's expertise relative to a task and their trust in AI shape AI-assisted work?} Our specific research questions are:

\begin{enumerate}[labelwidth=*,leftmargin=2.6em,align=left,label=RQ\arabic*]
 \item \label{rq:profile-outcome} How do \textbf{expertise and trust in AI} influence AI-assisted \textbf{work outcomes}, including task completion and human understanding?
 \item \label{rq:profile-interaction} What \textbf{interaction behaviors} explain how expertise and trust lead to different AI-assisted \textbf{work outcomes}?
\end{enumerate}

\begin{figure}[t]
  \centering
  \includegraphics[trim={0cm 5cm 3.5cm 0cm}, clip,width=\columnwidth]{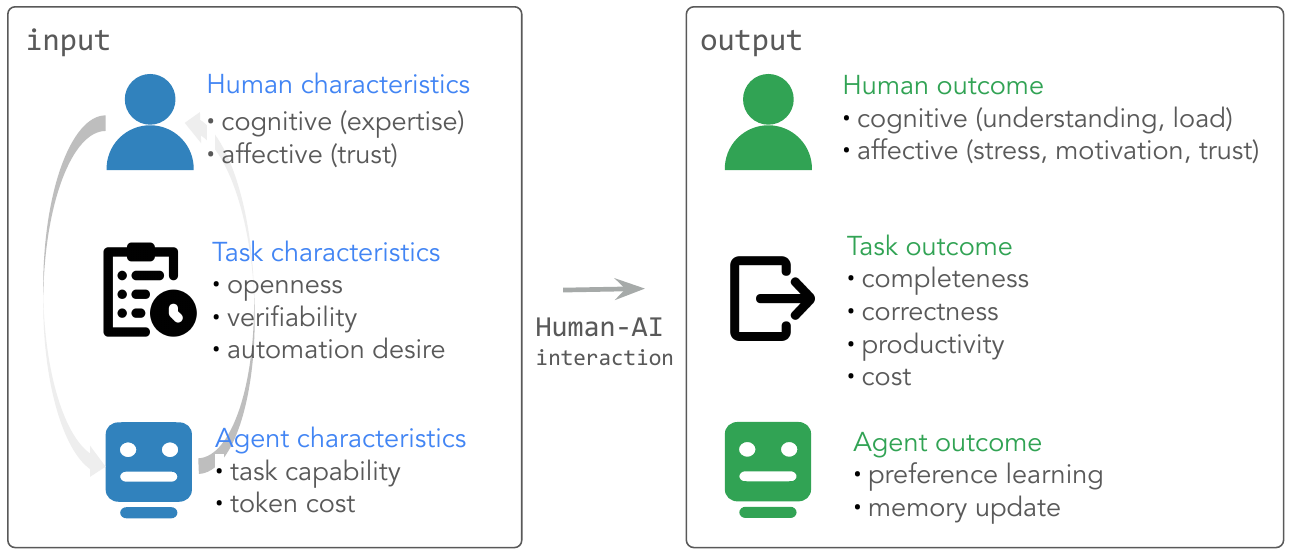}
  \caption{An interaction framework that describes how human, task, and agent characteristics influence the interaction and result in different human, task, and agent outcomes. }
  \label{fig:framework}
  \Description{Conceptual framework for studying AI-assisted work. Human characteristics, including cognitive factors such as expertise and affective factors such as trust; task characteristics, including openness, verifiability, and automation desire; and agent characteristics, including task capability and token cost, shape human-AI interaction. The interaction in turn produces human outcomes such as understanding and cognitive load, task outcomes such as completeness, correctness, productivity, and cost, and agent outcomes such as preference learning and memory updates.}
\end{figure}

Answering these questions poses a broader methodological challenge. AI-assisted work depends on \emph{who} the worker is, the task and agent they work with, \emph{how} they interact with the agent, and \emph{what outcomes} result. We organize these dimensions in a framework (Figure ~\ref{fig:framework}) for studying AI-assisted work. Viewed through this framework, common study paradigms leave different parts of this system unobserved or uncontrolled. Existing datasets of human--agent interactions~\cite{Baumann2026-rw, Mysore2025-bg} often capture trajectories and outputs, but not the human characteristics: we may know what someone did on a data-science task, for example, without knowing how much relevant expertise they brought to it. 
From these datasets, it is also difficult to isolate the effects of expertise and trust because tasks and agent setups may vary across users. 
Controlled experiments can hold these factors fixed, but often rely on short, bounded tasks rather than substantive end-to-end professional work~\cite{Bucinca2021-ii, Vasconcelos2022-mp}. Studying these relationships therefore requires a common, substantive task and agent environment in which workers with different profiles can be observed throughout a work cycle.

We address this challenge with an in-person think-aloud study in which 20 professionals --- designers, product managers, software engineers, and data scientists --- worked on a data-science take-home task previously used in hiring at a major U.S. software company. This design lets us define expertise relative to a common task, recruit participants spanning different levels of that expertise, and observe everyone working on the same problem with the same agentic tool. Crucially, the task is substantive enough that plausible-looking AI output can still contain fundamental errors, making both human steering and understanding consequential and observable. Specifically, the task asks participants to prototype an employee-attrition model using a dataset that deliberately lacks an attrition label, a limitation explicitly identified in the task requirement. Participants must construct a proxy label and reason about what the resulting analysis can legitimately demonstrate. By default (one-shot prompting), the AI agents introduce \textit{label leakage} where pipelines construct the proxy label using the same variables subsequently used to predict it, yielding high accuracy without meaningful evidence of attrition prediction. For each participant, we use a survey to capture characteristics such as task-relevant expertise, and collect full interaction logs, screen and think-aloud recordings, their task solution, and a post-task interview. Together, these data let us examine how workers’ profiles and interaction patterns relate to the completeness and validity of their task solutions, as well as to their demonstrated understanding of the work.

We found that experts more often decomposed the problem, investigated assumptions, and steered the agent, whereas novices more often delegated substantive decisions and struggled to evaluate its output. Greater involvement was associated with better understanding but less complete notebooks, highlighting a tension between developing understanding and producing a finished artifact within the available time. Trust was associated with confidence, engagement, satisfaction, and cognitive load, but not with artifact quality. Participants who trusted the agent more felt more confident and engaged in the task, reported greater satisfaction, and experienced lower cognitive load. However, their artifacts were not rated more highly. We also found that what people said about delegation often differed from what they actually did. Stated willingness to delegate was only weakly related to how involved people remained during the task, whereas observed delegation was closely associated with task completion and understanding.

Our results offer an empirical view of AI-assisted work among professionals completing a real-world task. We view our findings as hypothesis-generating and discuss their implications for model builders, end users, and assessment designers.
In summary, this paper makes the following contributions:

\begin{itemize}[labelwidth=*,leftmargin=1.3em,align=left, topsep=0em]
  \item \textbf{A holistic
  framework with hypothesis-generating think-aloud data.} We articulate a framework (\cref{fig:framework}) for studying AI-assisted work that jointly models human \emph{cognitive} and \emph{affective} factors, \emph{task properties}, the \emph{interaction loop}, and \emph{outcomes}. We instantiate it in a small dataset ($N=20$) from a controlled think-aloud study on a realistic agentic data-analysis task. 
  
  \item \textbf{An in-depth analysis of workflows by expertise and trust.} We provide mixed-methods (quantitative and qualitative) evidence on how expertise and trust jointly shape both the \emph{process} and the \emph{outcomes} of agentic work on a realistic task, identifying behaviors that predict success.
  
  \item \textbf{Interaction recipes, failure modes, and design implications.} We characterize the failure modes that distinguish workers by expertise and trust, and translate them into implications for design and practice for building and teaching people with varying levels of expertise to work with agents.
\end{itemize}

\section{Related Work}
\label{sec:relate}

We situate our work in three areas:  (1) expertise in AI-assisted work, (2) trust and delegation comfort, and (3) prompting behaviors and interaction trajectories that shape AI use. 
In short, prior work either characterizes \emph{behavior} without calibrated \emph{human} characteristics like expertise and trust, or measures task outcomes without capturing impacts on humans using realistic tasks. Our study is designed to join all of these aspects (as shown in \cref{fig:framework}).

\subsection{Expertise in AI-Assisted Work}
\label{subsec:expertise}

As LLMs continue to take on more complex work that requires different domain knowledge and skills, a lot of work has explored the relationship between expertise and effective use of AI in work.
In customer-support, consulting, and coding tasks, prior work has found that LLMs reduce task completion time \cite{paradis2025much}, increase quality \cite{Noy2023-jy, Dell-Acqua2026-pz}, and could serve as a \emph{leveler}, where less-experienced and lower-performing workers benefit more \cite{Peng2023-uf, Brynjolfsson2025-fu}. 
On the other hand, there are studies that show LLMs as an \emph{amplifier} for domain experts, which benefit experienced people more in terms of task performance \cite{Ma2026-cd} and productivity gains \cite{Daniotti2026-yp}.
Meanwhile, some studies found LLMs to be a \emph{inhibitor} for domain experts, showing that using AI slows down professionals in open-source development \cite{Becker2025-ho}, reduces quality in long-horizon tasks \cite{He2026-tp} and LLM-hard tasks \cite{Dell-Acqua2026-pz}, and increases maintenance and review burdens disproportionately on experts \cite{Xu2025-xh}. Interestingly, \citet{Becker2025-ho} also found a perceived-versus-objective productivity gap, as developers perceived a 20\% speedup, while AI actually led to a 20\% slowdown.

Prior work also surfaced potential moderator variables, such as time (e.g., limited time could make LLMs a leveler that compresses skill differences, versus unlimited time making LLMs an amplifier that benefits experts more \cite{Ma2026-cd}). However, these studies focus on task outcomes such as productivity and quality, leaving out the important impact that human-AI interaction has on humans with different levels of experience \cite{Ma2023-as}. Moreover, there is no unified framework that describes those studies' setup, task characteristics, user populations, interaction behaviors, and target outcomes to explain the different results.
Our work addresses this gap by examining how workers with different expertise interact with the same agent on the same realistic task, and explores different outcomes including task completion, validity, human understanding, and human perception, all described in a unified framework (\cref{fig:framework}).

\subsection{Trust and Delegation in AI-Assisted Work}
\label{subsec:profiling}

Beyond expertise, the impact of users' trust in AI is also widely explored, especially in AI-assisted decision-making literature \cite{Raees2026-zp}. 
Calibrated trust is important to human-AI collaboration, as overreliance can produce harmful outcomes, but underreliance can be unproductive \cite{Liu2026-ol}.
Some prior work showed that calibrating human self-confidence enhances human-AI team performance and encourages more rational reliance and aligned delegation on AI~\cite{shuai2024are, pinski2023aiknowledge}, and subjective trust in AI is a primary driver of delegation~\cite{Biswas2026-mp}. 
However, there is also empirical evidence that trust may not inform users' appropriate reliance on AI systems \cite{Lai2023-mv}, and reported trust and actual reliance can diverge \cite{Spillner2026-zy}. For example, a worker might trust that an agent is highly accurate yet still prefer not to delegate because the task is personally meaningful or high-stakes, or the worker values authorship~\cite{Lee2023-xb}. Recent work found that self-rated delegation comfort contributes to human–agent collaboration skill more than trust~\cite{Shao2026-hp}, but how these affective factors influence human-agent collaborative work outcomes is still unclear.

Much of the prior literature focused on exploring the relationship between trust (an affective attitude) and reliance (a behavior) \cite{Lee2004-yh} in decision-making tasks such as image classification or delegation tasks such as operating autonomous vehicles~\cite{wichnewski2023measuring}. 
Many interventions have also been developed, primarily targeting simple task outcomes (e.g., accuracy), affective outcomes (e.g., calibrated trust), or behavioral outcomes (e.g., calibrated reliance) \cite{Bucinca2025-lc, Raees2026-zp}.
As modern LLM-based agents are designed to be more autonomous and can perform much more complex tasks in knowledge work, how human trust impacts delegation and leads to diverse work outcomes needs further examination \cite{Mitchell2026-hx}. 

We aim to address this gap by exploring the role of trust and reliance in complicated AI-assisted knowledge work tasks using generative agents. 
In our study, by collecting human characteristics, interactions, and work outcomes, we aim to explore how factors like trust influence delegation behaviors and how that shapes the downstream task performance and human experiences.

\subsection{Prompting Behaviors and Human-Agent Interactions}
\label{subsec:behaviors}

A growing body of work analyzes \emph{how} people interact with LLMs across multi-turn interactions and how different interactions or prompting behaviors might impact task outcomes. 
From open-source programming data \cite{Baumann2026-rw, He2026-tp,Tang2026-ph}, writing in the wild \cite{Mysore2025-bg}, to specifically collected datasets from courses \cite{McNichols2026-uc, He2026-lz, Lucchetti2025-da}, quantitative user studies \cite{Mozannar2024-wq,Shin2025-ik}, and qualitative lab studies \cite{Mozannar2024-wh, Zamfirescu-Pereira2023-is}, many studies have attempted to answer the question: what interaction behaviors lead to \textit{good} outcomes? While most work operationalized good outcomes as task success, performance, and productivity~\cite{Shao2026-hp}, some explored the human aspects, such as improving learning \cite{Ma2025-wf,Varone2026-fj} and reducing overreliance~\cite{Liu2026-ol}.

These works often attempted to surface taxonomies to describe effective vs. ineffective behaviors with regard to their target outcomes. 
For example, defining the audience, communicating tone and style, and providing examples might contribute to session success~\cite{Shao2026-hp}; 
students asking for explanation or guidance during AI interaction (adaptive help-seeking) achieved higher exam grades than students who simply asked AI to produce the full solution \cite{Varone2026-fj}; and behavioral indicators such as careful task comprehension and fine-grained navigation lead to low overreliance \cite{Liu2026-ol}.
Together, prior work provides useful descriptions of interaction behavior, but few studies provide the full picture of how users with different characteristics can shape different interaction trajectories, and subsequently influence diverse work outcomes. 

We propose a framework (\cref{fig:framework}) to situate our work within it and more clearly describe the full story. 
We address this gap by connecting users’ cognitive and affective profiles to their interaction trajectories and to separate measures of task completion, artifact validity, and understanding.

\section{Method}
\label{sec:method}

To study how expertise and trust shape AI-assisted work, we need to jointly capture, under the same agent and task environment setup, \emph{who} the worker is, \emph{how} humans interact with the agent, and \emph{how good} the outcome is in terms of both task completion, validity, and human factors such as understanding and stress (\cref{fig:framework}).

Our framework  (\cref{fig:framework}) mirrors prior human-AI interaction models in the relevant factors such as human/AI and context/task characteristics \cite{McGrath2024-ix, Sun2023-du, Ma2023-as}. 
To organize different \textit{human characteristics}, we draw on educational and psychological taxonomies that distinguish cognitive and affective dimensions~\cite{Bloom1956-nv}. We measure domain expertise as \textit{cognitive} factors, and trust and comfort with delegation as \textit{affective} factors. We further distinguish characteristics measured before the task from task-dependent experiences, such as cognitive load, engagement, and understanding, which are \textit{human outcomes} (human cognitive and affective states) beyond task quality and efficiency. 

In this work, we conduct a controlled, in-person \emph{think-aloud study} (\cref{subsec:procedure}) with 20 professionals (\cref{subsec:participants}) that captures full end-to-end interaction trajectories together with rich context and outcome measures (\cref{subsec:measures}). This data uniquely joins human profiles (\cref{subsec:profiles}), interactions, and human and task outcomes, which enables rich quantitative and qualitative analysis to answer our research questions (\cref{sec:findings}).
Our think-aloud study was designed to explore how cognitive and affective human factors shape behavior and outcomes, and we leave large-scale confirmatory studies for future work (\cref{subsec:limitations}). 

\subsection{Study Procedure}
\label{subsec:procedure}

Each session lasted approximately two hours, one-on-one with an author, and proceeded through six segments: (1) a \emph{setup and onboarding} (${\sim}20$ min); (2) a \emph{pre-session survey} (${\sim}10$ min) capturing background, domain and AI-use skill, and affective measures; (3) the \emph{main task} (30 min), an AI-assisted data-analysis task; (4) two \emph{subtasks} (30 min) performed \emph{without} AI; (5) a \emph{post-session survey} (${\sim}10$ min) capturing cognitive load, understanding, engagement, confidence, and reflection; and (6) a \emph{debrief} (${\sim}10$ min) with a short semi-structured interview. 

For the main task, participants worked in VS Code with GitHub Copilot in \emph{agent mode}. To hold the model constant across participants, we fixed the underlying model to \texttt{claude-sonnet-4-6}.
The full log was exported from Copilot and uploaded with the artifacts by the participants at the end of the study.

\subsection{Task Design}

The main task was adopted from a real take-home assessment for a data scientist role from the organization, which allowed multiple days for interviewees without using AI. This is a task where neither a human nor a model alone can adequately complete it within 30 minutes, and we limit the time on task to simulate time-constrained scenarios that often happen in real work, and also to observe what workers will prioritize under time constraints. 

The main task asked participants to act on a realistic, intentionally \emph{under-specified} product request: design and build a Minimum Viable Product \emph{employee attrition model} that predicts whether an employee leaves the company or not using a workplace-analytics dataset\footnote{\texttt{pq\_data} from the \texttt{vivainsights} Python package: \url{https://microsoft.github.io/vivainsights-py/}. Full main task materials in \cref{app:main-task}.}
to inform whether the feature merits engineering investment.
Participants were asked to produce a Jupyter notebook containing their recommended modeling approach and rationale, end-to-end code, example outputs (predictions, feature importance, evaluation metrics), and a discussion of production considerations.\footnote{{Completeness spanned the full 0–100\% range (median 50\%), so the task discriminated without flooring or ceiling. 11 of 20 submitted notebooks (or out of 15 notebooks that present a trained model) contain label leakage — the task reliably produced the failure mode it was designed to elicit.
}} 

The task was deliberately designed to require \emph{handling ambiguity}: the dataset \textbf{lacks a real attrition label} (i.e., whether the employee left the company or not), so the dependent variable must be engineered or simulated. The task brief explicitly called out this dataset limitation, and we specifically look for participants' understanding of this key gap after task completion. We chose an open-ended, low-verifiability task precisely because such tasks are where human interaction and human judgment are expected to matter most. 

In addition, we verified in our pilot and task design phase that simply one-shot generating a solution given the task brief produces a \textbf{label-leakage failure}: the agent circularly synthesizes a proxy label out of the very features it then uses as predictors, so the model reports near-perfect accuracy ($\sim$0.90+) that is substantively meaningless. As a fluent, complete-looking notebook is the \emph{default} agent output and the leakage is hidden in the metrics, domain judgment and true understanding, e.g., how the label and features are constructed, separate a valid solution from a plausible artifact.

After the main task, participants completed two subtasks \emph{without} AI, to improve a plan and evaluate a notebook on a related but distinct analysis using the same dataset (adoption of a specific tool and its relationship to collaboration patterns)\footnote{Full subtask materials in Appendix \cref{app:subtask-context,app:subtask1,app:subtask2}.}. 
Making these subtasks human-only isolates evaluation and steering skill from generation, so that a participant's ability to fix a flawed plan or output is measured directly rather than confounded with the agent's contribution. 
The subtasks enable us to analyze what types of edits people with different expertise and trust make, starting from the same input plan and output artifact.

\subsection{Measures}
\label{subsec:measures}

As we fixed the task and agent used in our study, we collected four classes of measures corresponding to the framework (\cref{fig:framework}, summarized in
\cref{tab:measures}): human characteristics (from the pre-survey), interaction measures (from logs and think-aloud), task outcomes (from artifacts), and human outcomes (from post-survey and post-interview). Some survey measures were adopted from prior work \cite{Shao2026-hp, Ma2026-cd}, and full surveys are in \cref{app:pre-survey}.

\begin{table}[t]
\centering
\caption{The four classes of measures collected in the study mapping to the framework (\cref{fig:framework}), with their primary instruments and sources.}
\label{tab:measures}
\small
\begin{tabular}{@{}p{0.25\columnwidth}p{0.7\columnwidth}@{}}
\toprule
Construct & Instrument / source \\
\midrule
\emph{Human characteristics} & Pre-session survey: domain and AI-use skill, tool experience, task confidence; trust, self-efficacy, reliance, delegation comfort, etc. \\
\addlinespace
\emph{Interaction process} & Copilot prompt--response logs, screen recordings, think-aloud transcripts: turns, revisions, time per step, prompt behavioral codes (delegation vs.\ decomposition, help-seeking, verification/critique events, etc.). \\
\addlinespace
\emph{Task outcome} & Main-task notebook and subtask docs: artifacts scored on completeness and epistemic validity; number of weakness detected/corrected. \\
\addlinespace
\emph{Human outcome} & Post-session survey and interview: cognitive load, engagement, self-rated success, output-quality confidence, AI-use satisfaction, task understanding, etc. \\
\bottomrule
\end{tabular}
\Description{Summary of the four classes of measures collected in the study and their sources. Human characteristics come from the pre-session survey; interaction-process measures come from Copilot logs, screen recordings, and think-aloud transcripts; task outcomes come from graded notebooks and human-only subtasks; and human outcomes come from the post-session survey and interview.}
\end{table}

\paragraphBold{Human characteristics (cognitive and affective traits)}

Expertise is measured with \emph{self-report} Likert scales for domain/data-science skills, including years of experience, frequency of use in work, and task-specific confidence (summarized in \cref{subsec:profiles}).
The pre-survey also captures affective factors, i.e., reliance on AI, trust in AI outputs, comfort delegating (including high-stakes tasks), and beliefs about the appropriate level of automation (from human-led to AI-led). 

\paragraphBold{Interaction process}
From the logs, recordings, and think-aloud transcripts, we collect full prompt-response trajectories and interaction metadata (number of turns, revision count, time per step, agent plans and traces). We manually coded these trajectories. To surface the workflows and usage strategies that emerged, we inductively coded the trajectories. And separately, to classify the extent of observed delegation, we deductively coded each session according to the five levels of the human-agent delegation scale~\cite{Shao2025-ui, fereday2006demonstrating}.

\emph{Inductive: workflow characterization.}
To understand \emph{how} participants worked with agents, two authors open-coded the think-aloud transcripts, screen recordings, and prompt trajectories \cite{strauss1998basics}.
We fixed only the workflow phases (\emph{plan}, \emph{implement}, \emph{review}, \emph{outcome}) and let the strategy within each phase emerge from the sessions, yielding codes such as \emph{copy-paste} versus \emph{explore and decompose} at planning, and \emph{surface review} versus \emph{critical guidance} at review. The two authors coded all 20 sessions independently, reached strong agreement after constructing the codebook (Cohen's $\kappa=0.84$) \cite{McHugh2012-sc}, and then resolved every disagreement through discussion. Following common practice for interpretive qualitative analysis, we report agreement as an indicator of coding stability and do \emph{not} treat these codes as measurements to be tested statistically.

\emph{Deductive: observed delegation level (H1--H5).}
To understand how much humans actually delegate tasks to agents during the interaction, two authors deductively coded each session on a five-level human-agent delegation scale capturing how much of the decision the human retained \cite{Shao2025-ui, fereday2006demonstrating}. The annotation achieved substantial agreement (Cohen's quadratic-weighted {$\kappa=0.65$}), and disagreement was resolved via discussions. Participants with annotated H\,$\geq3$ (H1 autonomous agent → H5 full human involvement) are referred to as \emph{hands-on} and the rest \emph{hands-off} in this paper. The same H1–H5 scale is administered as a single self-report item pre- and post-session, which lets us compare \emph{stated} against \emph{observed} delegation on one scale (\cref{sec:rq2-saydo}).

\paragraphBold{Task outcome}

For the main-task notebook, as polished-looking notebooks are not always well-reasoned, we measure outcomes on two axes: \emph{completeness} and \emph{validity}. 
We score each notebook against a fixed rubric with explicit no-/partial-/full-credit anchors per item ($0/1/2$ points each).
\emph{Completeness} covers the task's four required outputs: a stated algorithm recommendation with rationale, executed end-to-end code, example outputs, and a production discussion (4 items, $/8$).
\emph{Validity} covers ambiguity and common mistakes in data processing (explicitly address the missing label), label construction (reasonably simulate attrition), modeling (avoid label leakage), and evaluation (avoid test set leakage) stages (4 items, $/8$). Any validity item may instead be marked \emph{N/A} when the condition it assesses never arose---for instance, model and evaluation cannot be assessed in a notebook that never trained a model, so validity is reported as a proportion of each notebook's applicable maximum.

Grading was performed by two members of the research team with data science expertise. To support this process, we provided the graders with an LLM-based grading aid (OpenAI \texttt{gpt-5.6-terra}) that produced a preliminary score, a justification, and citations to the notebook cells used as evidence. The two expert graders split the items and independently verified each preliminary score against the notebook, correcting it when necessary. Across the 160 item-level scores, the experts corrected 15 LLM-assigned scores. Most corrections involved \texttt{model\_validity}, where the LLM grader sometimes failed to identify label leakage. We use only the human-verified scores in all subsequent analyses of task outcomes. Full prompts and rubrics are reported in Appendix \ref{app:grading}.

For the human-only subtasks, we are interested in what the participant noticed in a flawed plan or output. To help us understand what seeded errors can be discovered by humans, an author graded all subtask submissions to annotate the flaws participants caught (full rubric in Appendix \ref{app:subtask-rubric}).

\paragraphBold{Human outcome}
The post-session survey captures both cognitive and affective outcomes, including task understanding, cognitive load (mental demand, effort, frustration, time pressure from NASA-TLX \cite{Hart1988-kv}), engagement, self-rated success, confidence in output quality, satisfaction with AI use, and self-rated task-context understanding. 

The post-interview and think-aloud also capture participants' understanding of the task, including whether participants recall the actual model developed in the task, what constitutes the proxy label, what constitutes the independent features, whether participants understand that there is a missing attrition label in the dataset which was simulated, and if there is a label leakage issue in the developed model. Participants' answers to these five understanding questions are clearly labeled as yes (2 pts) / partial (1 pt)/ no (0 pts), or unimplemented, e.g., if they have not reached the stage of label construction.

\subsection{Participants}
\label{subsec:participants}

We recruited 20 professionals (9 women, 11 men) at a large technology company, stratified by job role into four groups of five --- \emph{data scientists} (DS), \emph{designers} (Design), \emph{product managers} (PM), and \emph{software engineers} (SWE) --- with deliberate variation in job level (9 juniors, 11 seniors) so we have a wide coverage of professional experience (as summarized in \Cref{tab:participants}). Each participant received a unique identifier (P1-P20) and was compensated \$100 for their participation in the two-hour study. We note that this is a convenience sample drawn from a single organization and is not intended to be representative. We aimed to span a broad range of cognitive and affective profiles rather than to estimate population parameters.

\subsubsection{Human Profiles: Expertise and Trust}
\label{subsec:profiles}

    
\begin{table}[t]
\centering
\caption{The human-profile factors, mean inter-item correlation, Cronbach's $\alpha$, variance captured by the first principal component of constituent items, and inter-factor correlation.}
\label{tab:profiles}
\small
\begin{tabular}{@{}lrrrr@{}}
\toprule
Factor & Items & Mean $r$ & $\alpha$ & PC1 var. \\
\midrule
In-domain expertise    & 28 & 0.56 & 0.97 & 0.58 \\
AI-delegation trust     &  4 & 0.55 & 0.82 & 0.67 \\
\midrule
\multicolumn{5}{@{}l}{\emph{Inter-factor correlation}} \\
In-domain exp. $\times$ AI-deleg. trust & \multicolumn{4}{r}{$+0.51$} \\
\bottomrule
\end{tabular}
\Description{Summary of the two human-profile factors used in the analysis. In-domain expertise combines 28 task-relevant experience, frequency, and confidence items and has Cronbach's alpha of 0.97. AI-delegation trust combines four trust and delegation-comfort items and has an alpha of 0.82. The two factors have a moderate positive correlation of +0.51.}
\end{table}

From the pre-survey, we derive two interpretable \textit{profile factors}---cognitive (expertise) and affective (trust)---that serve as our independent variables. 
Specifically, we ordinally encode each survey item and $z$-score every item before mean aggregation. The scores for each participant are reported in \cref{tab:participants}. Guided by an item-level correlation and PCA screen of the full pre-survey matrix, we derive (\cref{tab:profiles}): 
\begin{itemize}[labelwidth=*,leftmargin=1.3em,align=left,topsep=0em]
  \item \textbf{In-domain expertise} (28 items, $\alpha=0.97$): years of experience, no-AI usage frequency, self-rated confidence, and AI-assisted usage frequency, each rated across seven data-science skill areas (EDA, statistical analysis, ML/model building, visualization, working with messy data, deriving business insight, and communicating to non-technical audiences). This captures domain skills exercised, with and without AI.
  \item \textbf{AI-delegation trust} (4 items, $\alpha=0.82$): trust in AI for data-science tasks, trust in AI without human oversight, comfort delegating data-science tasks, and comfort delegating high-stakes tasks to an agent.
\end{itemize}

\begin{figure}[t]
  \centering
  \includegraphics[trim={0cm 0cm 0cm 0cm}, clip,width=\linewidth]{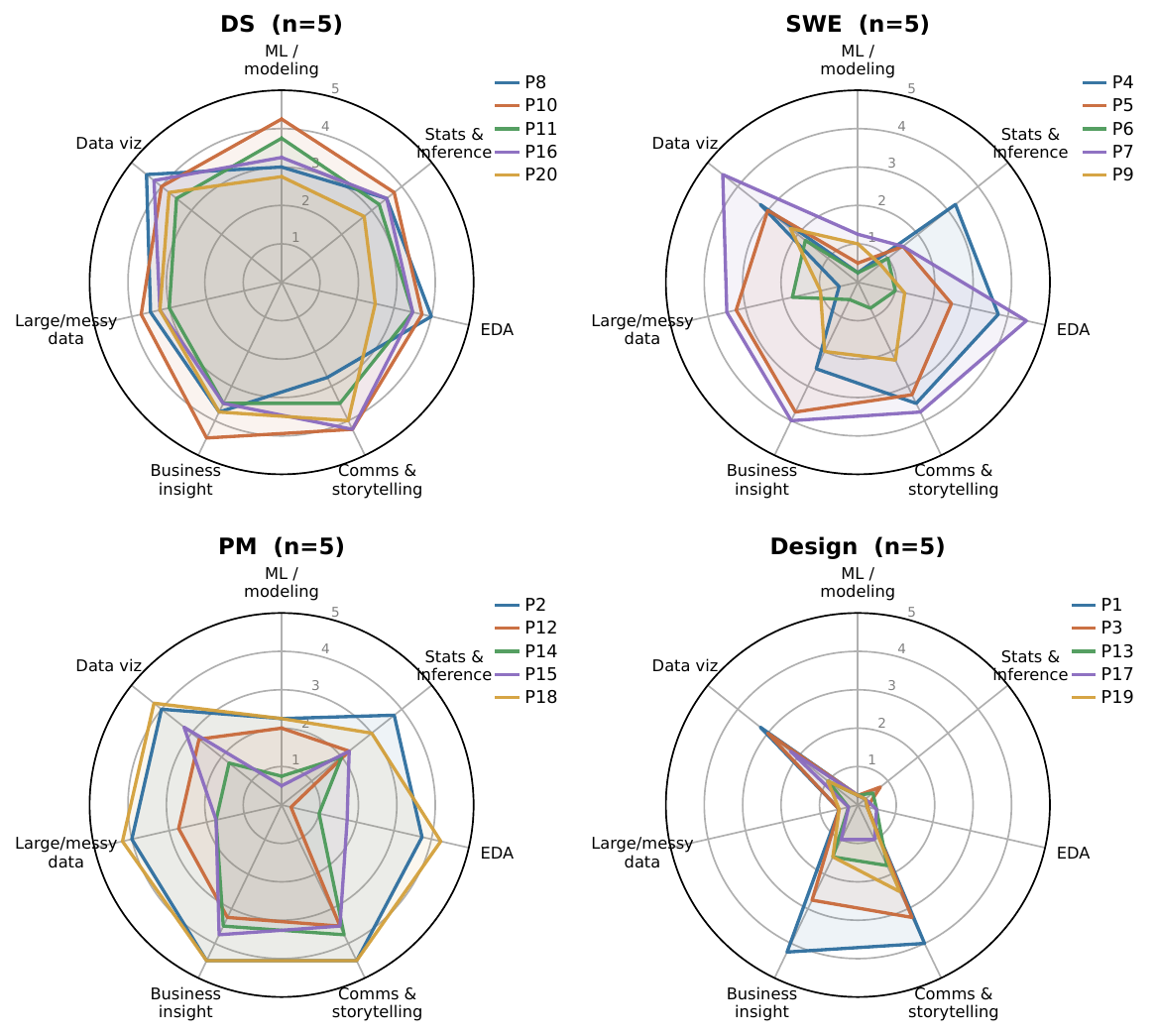}
  \caption{Self-rated in-domain expertise profiles by job role. }
  \label{fig:profiles-radar}
  \Description{Four radar charts showing participants' self-rated expertise across seven data-science skill areas, grouped by job role: data scientist, software engineer, product manager, and designer. Data scientists generally show broader and higher expertise across machine learning, statistics, exploratory data analysis, communication, business insight, large or messy data, and visualization, while participants in the other roles show lower or more uneven task-relevant expertise profiles. SWEs lead on visualization but not ML; PMs lead on insight and communication; designers sit lowest on the technical axes.}
\end{figure}

\begin{table*}[t]
\centering
\caption{Participant sample ($N=20$), four roles $\times$ five each. Yrs.\ = years in current role. \emph{In-domain expertise} and \emph{AI-delegation trust} are $z$-scored profile factors (\cref{subsec:profiles}). Expertise\,$\times$\,trust profile (Expert: expertise $\ge +0.31$; Trusting: trust $\ge -0.43$). }
\label{tab:participants}
\vspace{-5pt}
\small
\begin{tabular}{@{}llllllrrrc@{}}
\toprule
ID & Age & Gender & Ethnicity & Role & Level & Yrs. & In-domain exp. & AI-deleg.\ trust & Profile \\
\midrule
P1 & 35--44 & W & White & Design & Senior & 5--10 & $-0.37$ & $-0.96$ & Skeptical Novice \\
P2 & 35--44 & M & Multiracial & PM & Senior & 5--10 & $+0.95$ & $+1.62$ & Trusting Expert \\
P3 & 35--44 & W & Asian & Design & Junior & 5--10 & $-0.57$ & $-0.77$ & Skeptical Novice \\
P4 & 35--44 & M & White & SWE & Senior & $>$10 & $-0.06$ & $-1.30$ & Skeptical Novice \\
P5 & 25--34 & W & Hispanic & SWE & Junior & 3--5 & $+0.06$ & $+0.25$ & Trusting Novice \\
P6 & 25--34 & W & Asian & SWE & Junior & 0--1 & $-1.08$ & $-0.67$ & Skeptical Novice \\
P7 & 35--44 & W & Asian & SWE & Senior & $>$10 & $+0.56$ & $+0.63$ & Trusting Expert \\
P8 & 35--44 & M & White & DS & Senior & $>$10 & $+0.71$ & $-0.77$ & Skeptical Expert \\
P9 & 25--34 & W & Asian & SWE & Junior & 3--5 & $-0.71$ & $-0.08$ & Trusting Novice \\
P10 & 45--54 & M & White & DS & Senior & $>$10 & $+1.08$ & $+0.06$ & Trusting Expert \\
P11 & 25--34 & M & Asian & DS & Junior & 3--5 & $+0.71$ & $+1.20$ & Trusting Expert \\
P12 & 25--34 & W & Asian & PM & Senior & 5--10 & $-0.03$ & $+0.04$ & Trusting Novice \\
P13 & 25--34 & M & White & Design & Senior & 5--10 & $-1.16$ & $-0.04$ & Trusting Novice \\
P14 & 25--34 & W & Asian & PM & Junior & 3--5 & $-0.24$ & $+0.34$ & Trusting Novice \\
P15 & 25--34 & M & Asian & PM & Junior & 3--5 & $+0.03$ & $+0.86$ & Trusting Novice \\
P16 & 35--44 & M & Hispanic & DS & Senior & $>$10 & $+0.83$ & $-0.04$ & Trusting Expert \\
P17 & 35--44 & M & White & Design & Senior & 5--10 & $-1.18$ & $+0.25$ & Trusting Novice \\
P18 & 35--44 & M & Asian & PM & Senior & 3--5 & $+1.01$ & $+1.11$ & Trusting Expert \\
P19 & 35--44 & W & Asian & Design & Junior & 5--10 & $-1.11$ & $-1.52$ & Skeptical Novice \\
P20 & 25--34 & M & Asian & DS & Junior & 5--10 & $+0.57$ & $-0.18$ & Trusting Expert \\
\bottomrule
\end{tabular}
\Description{Participant-level summary for the 20 professionals in the study, including age range, gender, ethnicity, job role, seniority, years in role, standardized in-domain expertise, standardized AI-delegation trust, and descriptive expertise-by-trust profile. The sample contains five participants each from data science, software engineering, product management, and design and spans a wide range of expertise and trust.}
\end{table*}

\paragraphBold{Expertise profiles by role}
Our sample spans a wide range of \textit{competency shapes}: participants differ not only in how much in-domain expertise they have overall, but also in where that expertise lies (\cref{fig:profiles-radar}). We characterize expertise across seven task-relevant skill areas adapted from prior work~\cite{Ma2026-cd}: ML/modeling, statistical inference, EDA, communication, business insight, working with messy data, and data visualization. For each skill, we combine participants’ years of experience, frequency of working without AI, self-rated confidence, and frequency of AI-assisted work into a $0$--$5$ score. The resulting profiles show clear differences across roles: as we would expect, data scientists tend to have relatively broad expertise, while other roles show more uneven profiles with sharper strengths in particular areas. Overall expertise also varies considerably, with participant means ranging from $0.8$ to $4.0$/5.

\begin{figure}[bt]
  \centering
  \includegraphics[width=.8\linewidth]{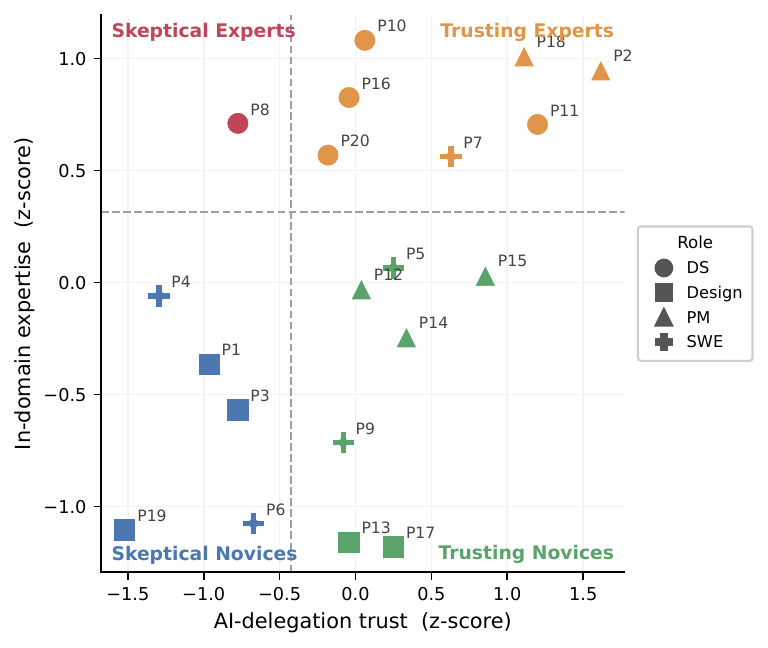}
  \caption{Participants' profiles in the expertise\,$\times$\,trust plane. Axes are mean $z$-scored \emph{in-domain expertise} (vertical) and \emph{AI-delegation trust} (horizontal).}
  \label{fig:profiles2x2}
  \Description{Scatter plot of the 20 participants by in-domain expertise on the vertical axis and AI-delegation trust on the horizontal axis. Participants are labeled by ID and shaped by job role. Dashed thresholds divide the plot into trusting experts, skeptical experts, trusting novices, and skeptical novices. The sample contains seven trusting experts, one skeptical expert, seven trusting novices, and five skeptical novices.}
\end{figure}

\paragraphBold{Expertise\,$\times$\,trust profiles}
Finally, we group participants into four profiles using the expertise and trust scores (\cref{fig:profiles2x2}). We cut each factor at its widest natural break (the largest gap in the sorted scores). This separates high- from low-skill professionals 
(gap ${\approx}0.50$ at $z{\approx}{+}0.31$) 
and high- from low-trust users 
(gap ${\approx}0.49$ at $z{\approx}{-}0.43$). 

Crossing the two yields \emph{trusting experts} (high expertise, high trust; $n=7$), \emph{skeptical experts} (high expertise, low trust; $n=1$), \emph{trusting novices} (low expertise, high trust; $n=7$), and \emph{skeptical novices} (low expertise, low trust; $n=5$). Note that these labels are descriptive groupings used to structure qualitative contrasts, not a claim of discrete latent clusters.

Although recruiting participants by role helped ensure our participants spanned a wide range of expertise, it is a coarse characterization of an individual's actual expertise. For example, a particular PM may have a data science background, or a SWE may primarily focus on data analysis in their day-to-day work. Thus, in the rest of this paper, \emph{when we refer to experts vs. novices, we use the pre-survey in-domain expertise z-score split} instead of the job roles (data scientist vs. the rest).

\subsection{Data Analysis}
\label{subsec:analysis}

{Because both completeness and validity task outcome axes are bounded sums of $0/1/2$ items with pronounced floor and ceiling clustering, and Shapiro--Wilk rejects normality, we use non-parametric rank methods throughout: Spearman's $\rho$ for association, Mann--Whitney $U$ for two-group comparisons, and reported with the rank-biserial correlation $r$ as the effect size~\cite{Berry2018-vl}.

Three notebooks scored $0/8$ on completeness are dropped from validity analysis (3/4 or 4/4 NA validity items), which measures the absence of an artifact rather than poor reasoning. Two of these participants did not get to plan the solution and are thus dropped from understanding analyses (4/5 NA understanding items). Therefore, the final samples are $N=20$ for completeness, $N=17$ for validity, and $N=18$ for understanding.

\section{Think Aloud Study Results}
\label{sec:findings}

To better understand the interplay between human profile, interaction process, and outcome variables (\cref{fig:framework}), we answer our two research questions in the following sections:
\ref{rq:profile-outcome} asks how a
participant's domain expertise and trust in AI shaped the human and task \emph{outcomes} (\cref{sec:rq1});
\ref{rq:profile-interaction} dives into answering how these human-profile factors influence the \emph{interactions}, characterizing different workers' workflows (\cref{sec:rq2}),
%
and explores how interaction patterns contribute to different outcomes, distilling practical recipes for working with agents (\cref{sec:rq3}).

\subsection{RQ1: Experts understand but may not finish; trust shapes the experience, not the artifact}
\label{sec:rq1}

\subsubsection{Expertise showed up in whether the work was right and understood, not whether it was finished}
\label{sec:rq1-diligence}

\begin{figure}[t]
  \centering
  \includegraphics[trim={0 0 0 0},clip, width=0.7\linewidth]{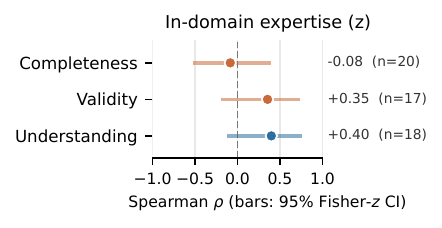}
  \vspace{-10pt}
  \caption{In-domain expertise's correlations with task completeness, validity, and human understanding.}
  \label{fig:expertise-outcome-corr}
  \Description{Forest plot showing Spearman correlations between in-domain expertise and three outcomes. Expertise is weakly negatively associated with artifact completeness, and weakly positively associated with artifact validity and human understanding.}
\end{figure}

When we graded the notebooks participants produced in terms of completeness and validity (\cref{fig:expertise-outcome-corr}), we observe that expertise is weakly positively associated with notebook validity and human understanding, and negatively associated with completeness; however, expertise is not clearly distinguishable from the notebook artifact. 

\begin{figure}[t]
  \centering
  \includegraphics[trim={0 5 5 0},clip, width=\linewidth]{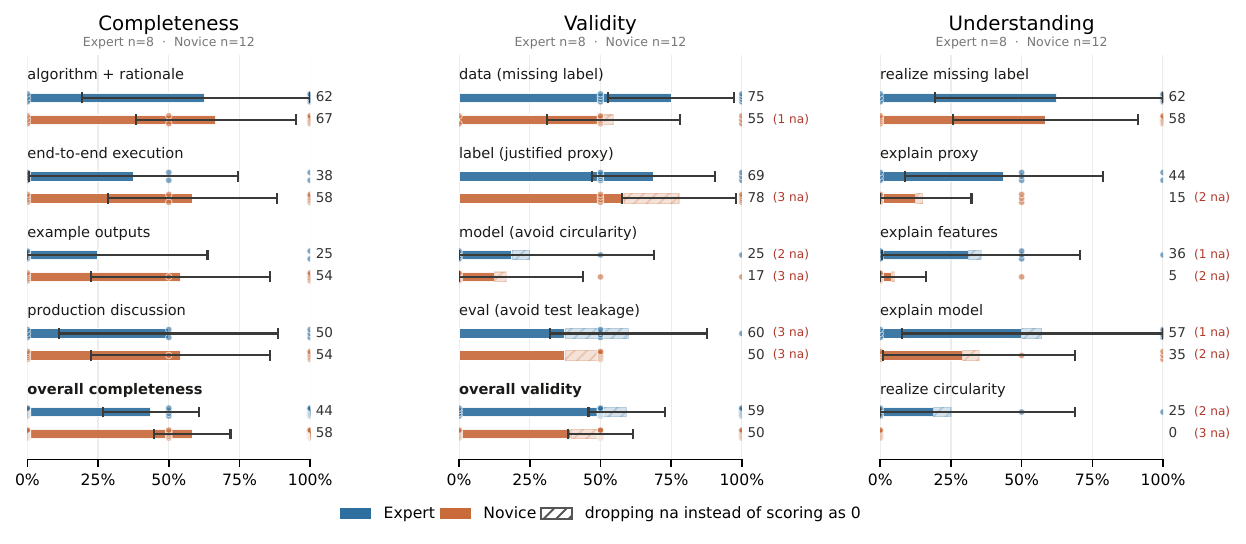}
  \caption{Detailed breakdown of experts' vs. novices' completeness, validity, and understanding scores by rubric items.}
  \label{fig:expertise-item-breakdown}
  \Description{Three grouped bar charts compare experts and novices on detailed rubric items for artifact completeness, artifact validity, and human understanding. Novices have higher overall completeness, especially for end-to-end execution and example outputs, while experts have somewhat higher overall validity and substantially stronger understanding of the proxy label, features, model, and circularity. Both groups frequently fail to recognize model circularity.}
\end{figure}

Further analyzing each rubric item (\cref{fig:expertise-item-breakdown}), we see that novices' notebooks are more complete than experts' (by 14.6\% on average), especially in terms of execution and outputs. 
Only 2 of 8 experts produced a fully executed pipeline and example outputs, compared to half of novices (6/12). 
Among the four notebook validity rubric items, the overall validity gap is 9\% (experts 59\% - novices 50\%), so experts present artifacts that appear more valid in data processing, model construction, and evaluation aspects.

In terms of human understanding, in-domain experts averaged more than twice the understanding of novices ($M=51\%$, $SD=44\%$ vs.\ $M=22\%$, $SD=21\%$; Cohen's $d=+0.86$). Realizing missing label is the least differentiating question, which is expected because this information is directly provided in the task brief. However, there are still 3/8 experts and 5/12 novices who could not answer whether the dataset has the attrition label or not.

\begin{figure*}[t]
\begin{subfigure}[t]{0.5\textwidth}
\centering
  \includegraphics[width=\linewidth]{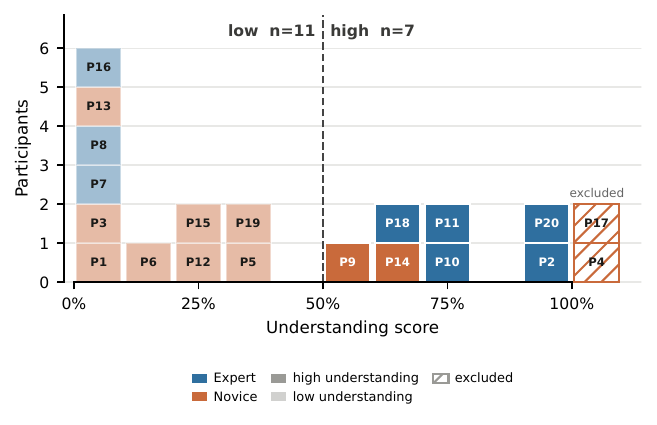}
  \caption{Understanding scores distribution. Two participants with 4/5 NA items are excluded from understanding analysis.}
  \label{fig:understanding-dist}
  \Description{A histogram of understanding scores from 0 to 100, where each participant appears as a labeled block stacked within their bin.}
\end{subfigure}%
\hfill
\begin{subfigure}[t]{0.48\textwidth}
\centering
  \includegraphics[trim={0 0 10 0}, clip,width=0.9\linewidth]{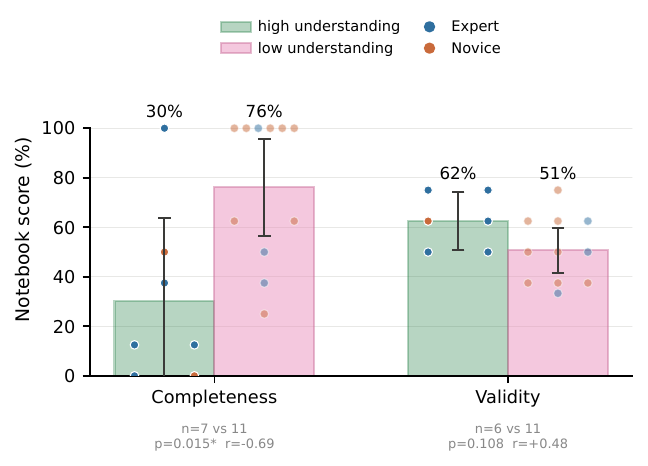}
  \caption{Different understanding by notebook completeness and validity, with Mann-Whitney U test $p$-value and rank-biserial $r$.}
  \label{fig:understanding-split}
  \Description{Paired bar chart.}
\end{subfigure}
\vspace{-10pt}
\caption{Tradeoff between human understanding and artifact completeness. High-understanding participants built significantly less complete notebooks, and somewhat more valid ones. }
\Description{Two-panel figure relating human understanding to artifact outcomes. Panel A shows the distribution of participants' understanding scores, with experts reaching the highest scores but several experts also scoring zero; novices are concentrated toward lower understanding. Panel B compares high- and low-understanding participants: the high-understanding group produces substantially less complete notebooks, about 30\% versus 76\%, but somewhat more valid notebooks, about 62\% versus 51\%. The completeness difference is statistically significant.}
\label{fig:understanding-coverage}
\end{figure*}

\paragraphBold{Expertise raised the ceiling on understanding, but not the floor} 

While some experts achieved the highest understanding scores in the study, three of eight scored zero (as shown in \cref{fig:understanding-dist}). Low-understanding experts were very hands-off when using AI; for example, 
P16 fully handed the work to three subagents and two loops, chose not to examine any intermediate thinking traces, and ran out of time before they could review the final output. 
In contrast, novices clustered at the lower end of the distribution, with none scoring above 60\%.
For example, no novice realized circularity in the model they implemented, while 2 experts did. Similarly, more than half of the experts (4/7) could explain the features their model used, while only 1 of 10 novices was able to do the same.

Taken together, 
in terms of the human understanding outcome, expertise seems necessary but not sufficient: it raised the ceiling on understanding without raising the floor.
Note that there can also be different types of understanding that the workflow influences differently, such as implementation understanding, dataset understanding, interpretation understanding, etc. We focus on implementation understanding in our study, and future work can explore more types of understandings' impact (\cref{subsec:limitations}). 

\paragraphBold{The diligence tax: completion and understanding are inversely stocked}
From the think-alouds, we can observe that human understanding and task completion are often two objectives produced by different workflows (\cref{sec:rq2-workflow}) and trade off with each other. 
A staggering case is novice designers P3 and P13, who achieved 100\% coverage in their notebook but walked away with 0\% post-task understanding, which means that they used AI to produce a complete notebook, but they did not realize the missing label in the dataset and could not explain how their model was constructed. 

Splitting participants at the midpoint of the understanding score yields seven \emph{high-understanding} and eleven \emph{low-understanding} participants (\cref{fig:understanding-dist}).
We can observe that high-understanding participants produced notebooks that were \textbf{less than half as complete} ($M=30.4\%, SD=36.0$) as those of low-understanding participants ($M=76.1\%, SD=29.3$); Mann-Whitney U $p=.015$, rank-biserial $r=-0.69$ (\cref{fig:understanding-split}). 
On validity, the difference runs the other way but is not significant (high-understanding $62\%$ > low-understanding \ $51\%$). 

\paragraphBold{Why understanding distinctly matters}
Participants themselves treated understanding as an outcome distinct from completion, for three reasons. The first is \emph{responsibility}: P13 separated the two axes explicitly --- \quoteinline{it's not very hard to complete the task. To complete the task well is a different question entirely} --- and P17 tied it to stakes, \quoteinline{if a task is not well-defined and I need to present the result to leadership and be responsible for it, I would not just one-shot it, in which case I won't understand what that AI did}. P14 described their deliberate counter-effort, \quoteinline{I've also been trying to keep myself honest and research what I don't know so I can own my work output}. 

The second is the downstream \textit{cognitive cost} of the absence of understanding, such as in comprehension and debugging. P9 named the mechanism by which comprehension and debugging fail under delegation: \quoteinline{if it generates a lot of code at once, then it becomes difficult for me to go through it and understand, not to mention fix its mistakes}. P20 similarly named the cost of not having built the notebook themselves: \quoteinline{when you're doing it yourself from the get-go, it's much easier\ldots If you use AI to build the whole process, then it's difficult for me to go back. It's actually a friction\ldots If I had done the feature exploration myself, this wouldn't overwhelm me as much, because I would immediately know what I need to look at.} 

The third reason is to assist in \textit{generating ideas}. P20 --- an expert data scientist achieving 100\% understanding and yet only at $12.5\%$ notebook completeness --- articulated it most fully. When asked what they would do with more time, they described  \quoteinline{building a mental map ... that helps you arrive at ideas, improve the final evaluation metric ... and think out-of-the-box ... If you start with the AI, you keep relying on it. There can be selection bias, as I am more involved in what the AI has produced now as compared to my original thought.}  The diligence tax and its potential compensating benefit are both visible in this one person.

\subsubsection{AI-delegation trust shaped human experience, not the artifact}
\label{sec:rq1-trust}

\begin{figure}[t]
  \centering
    \includegraphics[trim={0 0.2cm 6.5cm 0},clip, width=0.9\linewidth]{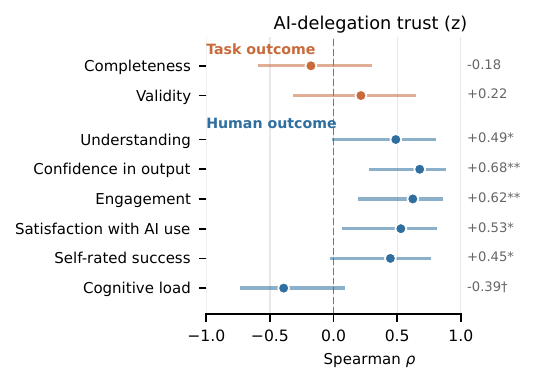}
  \vspace{-10pt}
  \caption{AI-delegation trust's correlations with task completeness, validity, human understanding, and human affective outcomes.}
  \label{fig:trust-forest}
  \Description{Forest plot showing associations between pre-task AI-delegation trust and task and human outcomes. Trust has little relationship with graded artifact completeness or validity, but higher trust is associated with greater human understanding, confidence in the output, engagement, satisfaction with AI use, and self-rated success, as well as lower cognitive load. The strongest association is with confidence in output quality.}
  \vspace{-10pt}
\end{figure}

As shown in \cref{fig:trust-forest}, AI-delegation trust measured before the session (\cref{subsec:profiles}) significantly correlates with \emph{human} understanding and affective outcomes, but not with \emph{graded} completeness nor validity task outcomes. Participants who reported more comfort delegating finished the session markedly more confident in an output whose measured validity or completeness was no better: confidence in output $\rho=+0.68^{**}$ ($p=.001$), engagement $+0.62^{**}$ ($p=.003$), satisfaction with AI use $+0.53^{*}$ ($p=.017$), self-rated success $+0.45^{*}$ ($p=.048$), and cognitive load $-0.39^{\dagger}$ ($p=.087$).

When participants do not trust AI, they may suffer from distress when they need to rely on AI for its judgment because of the lack of domain expertise themselves. For example, P19, a skeptical novice (designer), reported the least delegation comfort in the study (trust $z = -1.52$). Even though they were able to deliver a 63\% complete notebook, P19 reported the worst affective outcome we recorded --- high cognitive load of 4/5, low perceived success 1/5, low confidence 1/5, and low perceived understanding 1/5. During the interaction, there were also moments of voiced concerns and frustrations about needing to rely on AI judgment: \textquote[P19]{There is so much that I don't know! I don't know how to do this and so I feel uncomfortable asking AI to do it. But I need to.}.

\begin{figure*}[htbp]
  \centering
  \vspace{-15pt}
  \includegraphics[trim={0 2cm 0 0},clip,width=\linewidth]{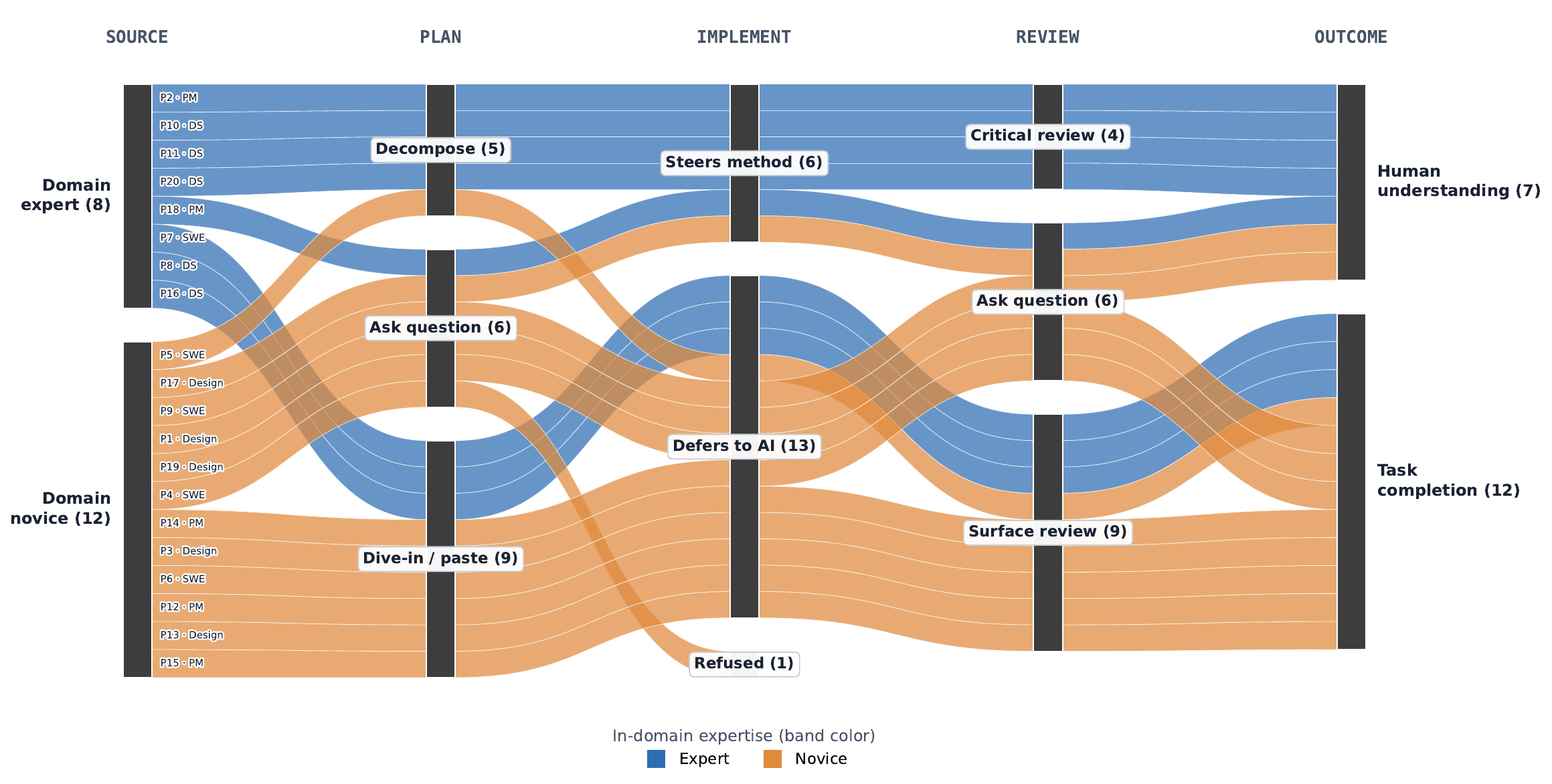}
  \caption{Workflow paths from source to outcome, colored by in-domain expertise. Experts concentrate in the decompose, steer, and critical-review paths that end in human understanding; novices concentrate in the dive-in/paste, defer-to-AI, and surface-review paths that end in task completion.}
  \label{fig:sankey-expertise}
  \Description{Sankey diagram showing participants' workflow paths through planning, implementation, review, and outcome, with paths colored by expertise. Experts are concentrated in explore-and-decompose planning, manual steering, and critical review pathways that more often end in human understanding. Novices are concentrated in dive-in or copy-paste planning, delegation to AI, and surface review pathways that more often end in task completion.}
\end{figure*}

\subsection{RQ2: Delegation behaviors led to completion workflows but traded human understanding}
\label{sec:rq2}

\subsubsection{Experts decompose and steer to produce understanding; novices paste and accept for coverage}
\label{sec:rq2-workflow}

Open-coding the think-aloud transcripts and screen recordings (\cref{subsec:measures}) yields a strategy per workflow phase for each participant (\cref{fig:sankey-expertise}; \cref{tab:strategy-outcome}). Expertise concentrates at particular points in the pipeline: four participants who \emph{steered the method manually} and gave \emph{critical guidance} at review are in-domain experts, and they are the same four people (P2, P10, P11, P20); five of the seven sessions ending in epistemic understanding are experts.
\emph{Where the decision lives seems to differ by expertise: experts decide, novices approve.}

Qualitatively, experts front-loaded exploration to build their own model of the data. P2 (explore-and-decompose) narrated \quoteinline{that's something I would want to validate to make sure I'm understanding what its EDA is telling me} and \quoteinline{I want to understand the data better\ldots I am kind of old school, so I want to visualize the data}. Novices more often delegated the framing itself: P15 explained \quoteinline{since I wanted to just learn about it, I just gave it intro information. Now that it's going through the full loop, I will let it complete its chain of thought}, adding that \quoteinline{if I had an opinion based on my prior learnings, I would have given it a lot more inputs}. A recurring novice substitute for review was to have the agent check itself --- P6 had {agent critique each other}, and P5 spawned {research, ML engineer, and people analytics subagents}.

\subsubsection{Stated delegation does not align with observed delegation in new tasks, and novices overclaim human involvement}
\label{sec:rq2-saydo}

Participants' stated and observed delegation were measured on the same five-point involvement scale (H1 autonomous to H5 full human involvement): once as a pre-session self-report, and once as a human annotation of the session itself (\cref{subsec:measures}). \textbf{The two do not align} (\cref{fig:say-vs-do}), and this dissociation between stated and observed delegation is also captured in prior literature comparing trusting attitudes and reliance behaviors \cite{Spillner2026-zy}.

The gap mostly occurs among novices. Novices ($n=12$) show a mean gap of $+1.08$, with seven overclaiming and \textbf{none} underclaiming; experts ($n=8$) show a mean gap of $-0.12$ (three over, five under). On this scale, \emph{novices seem to overestimate how much they will engage themselves, while experts are calibrated or underestimate}\footnote{Doing the task did not recalibrate the belief: only four of twenty self-ratings moved between pre- and post-session (P5, P7, P14, P18; mean $3.05 \rightarrow 3.15$), three moved up.}. 

\begin{figure*}[t]
  \centering
  \includegraphics[trim={0 0 0 21}, clip,width=.85\linewidth]{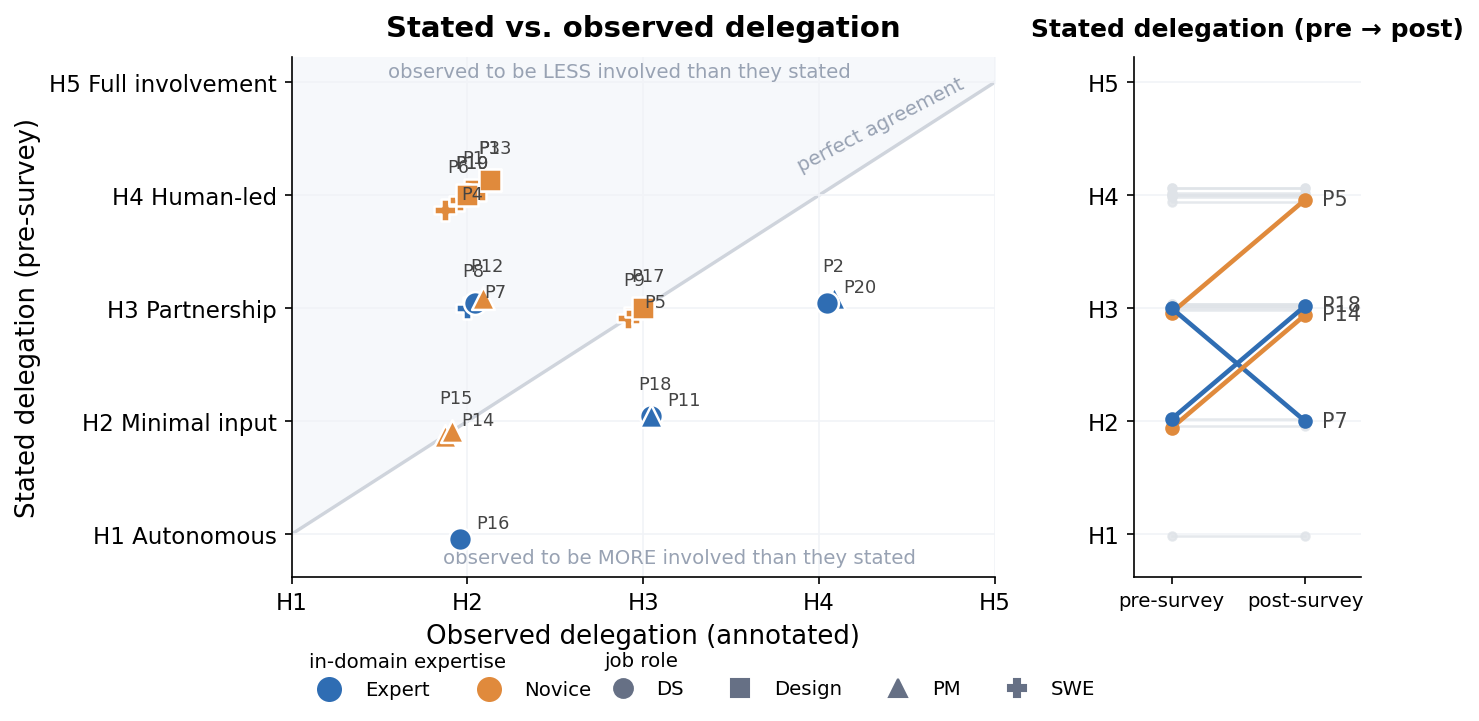}
  \caption{Stated versus observed delegation. Points above the diagonal are participants who overclaimed engagement (said they would be more involved than they were). The right panel presents the stated delegation before and after the session.}
  \label{fig:say-vs-do}
  \Description{Comparison of stated and observed delegation on a five-level scale from H1, autonomous AI, to H5, full human involvement. The left panel plots each participant's pre-session stated level against the level observed during the task; many novices are less involved in the unfamiliar task than their stated typical-work preference would suggest. The right panel shows that participants' stated delegation preferences change little between the pre- and post-session surveys.}
  \vspace{-10pt}
\end{figure*}

Note that the stated delegation was rated corresponding to participants' usual work, which could be different from the unfamiliar task setup. 
Some participants attribute their behavioral difference to the task. For example, P11 (a trusting but hands-on expert data scientist) described \quoteinline{in daily work [we] become the PM now, instead of looking at the coding files, coding outputs, we just look at the summaries and markdowns}, while noting that when given an unfamiliar dataset, they would start more manually to understand the task. 
In comparison, P16 was a trusting expert who chose to stay consistently hands-off in our task: \quoteinline{I don't verify thousands of lines of code. I will show you how I usually work.}

P13, a domain novice (senior designer) who self-rated as H4 (human-led) but was observed as H2 (AI-led), only sent two prompts during the whole session: ``...Please help me put a plan together in order to fulfill this request: [copied\_full\_task\_brief]'' and ``Does this MVP prototype include all of these requirements? [copied\_required \_output]''. P13 consistently voiced struggles to engage and provide instructions on their own:  \quoteinline{as I don't know ... I will let AI do it.}
P17, also a domain novice (senior designer), stayed H3 consistently during the study. While P17 spent the entire session trying to understand rather than blindly build, they still voiced the struggle to keep human engagement: \quoteinline{there is going to be a certain point where I just have to trust it and let it go}.

\begin{figure}[t]
  \centering
  \includegraphics[trim={0 0 0 0},clip, width=\linewidth]{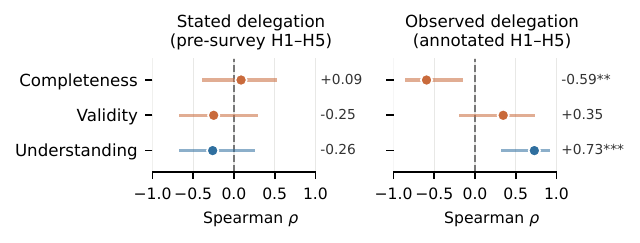}
  \caption{Stated delegation vs. observed delegation (H1 autonomous $\rightarrow$ H5 full human involvement) correlations with task completeness, validity, and human understanding.} 
  \label{fig:trust-engagement-corr}
  \Description{Two forest plots compare correlations of stated delegation and observed delegation with completeness, validity, and human understanding. Stated delegation has weak relationships with all three outcomes. In contrast, greater observed human involvement is strongly associated with lower completeness, rho equals -0.59, and higher understanding, rho equals +0.73, with a smaller positive association with validity.}
\end{figure}

Comparing the stated and observed delegation from interaction behaviors (\cref{fig:trust-engagement-corr}), observed delegation does significantly negatively correlate with completeness ($\rho=-0.59^{**}$, $p=.006$) but positively correlates with understanding ($\rho=+0.73^{***}$, $p=.001$). However, unlike AI-delegation trust (\cref{fig:trust-forest}), observed delegation does not correlate significantly with any affective outcomes. So the more human-led the interaction, the less complete the notebook, yet better human understanding as an outcome. 
For example, the hands-off novice P13 optimized completion ($100\%$ completeness, understanding $0\%$) by copy-pasting and reviewing at the surface. The hands-on novice P17 aimed for understanding, spending the session asking questions and ending with almost nothing built ($0\%$ completeness, but their only answerable 1/5 understanding item on missing label was correct). 


\subsection{Interaction recipes for different targeting outcomes}
\label{sec:rq3}

\begin{table*}[t]
\centering
\caption{Workflow strategy by phase, with the participants who used it, how many were in-domain experts, and the mean task outcome on each axis. Expert-only strategies are shaded.}
\label{tab:strategy-outcome}
\small
\setlength{\tabcolsep}{4pt}
\begin{tabular}{@{}r|l|p{0.40\linewidth}|rrr@{}}
\toprule
Phase & Strategy & Participants & \% Experts & Complete. & Validity \\
\midrule
\multirow{3}{*}{Plan}
 & copy-paste & P3, P6, P7, P8, P12, P13, P14, P15, P16 & 33\% & \textbf{82\%} & 51\% \\
 & ask question & P1, P4, P9, P17, P18, P19 & 17\% & 38\% & \textbf{50\%} \\
 & explore \& decompose & P2, P5, P10, P11, P20 & \textbf{80\%} & {18\%} & \textbf{65\%} \\
\midrule
\multirow{3}{*}{Implement}
 & delegate to AI & P1, P3, P5, P6, P7, P8, P9, P12, P13, P14, P15, P16, P19 & 23\% & \textbf{68\%} & 52\% \\
 & manual steering & P2, P10, P11, P17, P18, P20 & \textbf{83\%} & {27\%} & \textbf{62\%} \\
 & refuse to implement & P4 & 0\% & 0\% & --- \\
\midrule
\multirow{3}{*}{Review}
 & surface review & P3, P5, P6, P7, P8, P12, P13, P15, P16 & 33\% & \textbf{79\%} & 52\% \\
 & ask question & P1, P9, P14, P17, P18, P19 & 17\% & 46\% & \textbf{53\%} \\
 & \cellcolor{gray!12}critical guidance & \cellcolor{gray!12}P2, P10, P11, P20 & \cellcolor{gray!12}\textbf{100\%} & \cellcolor{gray!12}{16\%} & \cellcolor{gray!12}\textbf{62\%} \\
\midrule
\multirow{2}{*}{Outcome}
 & task completion & P1, P3, P5, P6, P7, P8, P12, P13, P14, P15, P16, P19 & 25\% & \textbf{74\%} & 52\% \\
 & human understanding & P2, P9, P10, P11, P17, P18, P20 & \textbf{71\%} & {23\%} & \textbf{62\%} \\
\bottomrule
\end{tabular}
\Description{Workflow strategies observed during planning, implementation, review, and final outcome, listing the participants using each strategy, the percentage who are domain experts, and mean artifact completeness and validity. Copy-paste planning, delegation to AI, and surface review produce relatively high completeness, whereas explore-and-decompose planning, manual steering, and critical guidance are disproportionately used by experts and produce higher validity but lower completeness.}
\end{table*}

Because there are independent dimensions of good work outcomes (\cref{fig:framework}), there is no single good strategy for AI-assisted work --- there is a strategy per target. Expertise and trust do not simply raise or lower an average outcome; they produce \emph{qualitatively different} ways of failing (\cref{sec:rq1}, \ref{sec:rq2}). Here, we summarize the interaction recipes that our study surfaces, and in each case name both the move that reaches desired outcomes and failure modes. 

\subsubsection{If the target is completion under time pressure: delegate, and know what you traded}
\label{sec:rq3-completion}

The completion recipe is legible in the strategy table (\cref{tab:strategy-outcome}): copy-paste the brief into the plan (achieving $82\%$ completeness in the final artifact), delegate implementation to AI ($68\%$), surface review ($79\%$), finish with a task-completion outcome ($74\%$). Overall, the more human involvement in the observed delegation scale, the less completed the notebooks are ($\rho=-0.59^{**}$; \cref{fig:trust-engagement-corr}). 

Within the completion recipe, the easiest verification move we observed is to \textbf{ask the agent to critique itself}. P6 had the agents critique each other, and P13 explicitly prompted the agent to check against the required output: `Does this MVP prototype include all of these requirements?''
Both P6 and P13 reached 100\% task completion. However, \emph{neither caught the label leakage in their model}, again demonstrating how task completeness trades with human understanding (\cref{sec:rq1-diligence}). 

{Another failure mode while prioritizing completion is that people may miscalibrate AI effort and overkill the task by spawning too many agents. For example, for both P16 expert and P5 novice who spawned subagents, the subagents took longer than they expected and resulted in them being unable to complete in time.}

\subsubsection{If the target is validity: domain judgment or careful review needed, reading does not suffice}
\label{sec:rq3-validity}

As shown in \cref{tab:strategy-outcome}, the moves that work for validity are explore \& decompose at planning (achieving $65\%$ validity in the final artifact), manual steering at implementation ($62\%$), and critical guidance at review ($62\%$).

Note that reading behaviors, although showing engagement, is not critical interrogation and may create a false sense of fluency and leave mistakes or misunderstandings undiscovered in surface-level reviews, as also highlighted in prior research \cite{Chi2014-ce}. In our study, P19 narrated \quoteinline{I don't understand it and cannot really pick out issues, but I will keep reading}. P3 also tried to read line-by-line and narrated most of the agent response. Both P19 and P3 ended with a low validity score of 38\%.

When the target is validity, people can fail to apply judgments: novices may not be able to apply domain-related supervision, but even experts may miss nuanced model errors when they need careful checking.
For example, {11 of the 15 submitted notebooks with implemented models contained label leakage, and none of those authors (4 experts, 7 novices) noticed}. Only two participants noticed or explicitly avoided leakage at all in their understanding questions, P20 and P10: both are domain experts and data scientists themselves. 
This label leakage failure is hard to capture, as it needs true understanding of the input and output variables, or noticing the superficially high model accuracy and carefully checking the label and feature construction.

\begin{figure}[t]
  \centering
  \includegraphics[trim={0 1cm 1.5cm 0.5cm},clip,width=\linewidth]{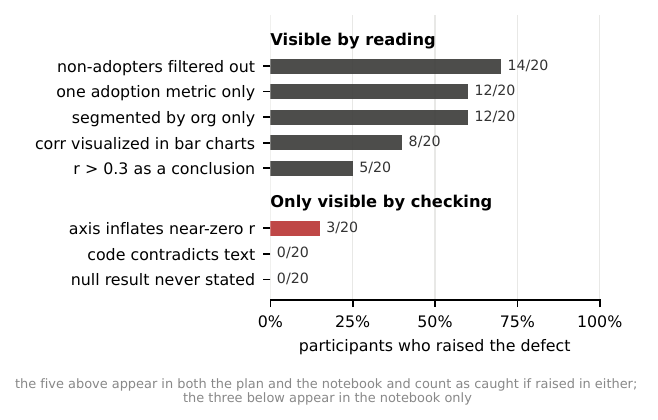}
  \caption{Catch rate by defect type in human-only subtasks.}
  \label{fig:defect-catch}
  \Description{Horizontal bar chart of catch rates, grouped into defects visible in the text, which range from 25 to 70 percent, and defects checkable only against the evidence, which range from 0 to 15 percent.}
\vspace{-10pt}
\end{figure}

This failure detection behavior reproduces in the human-only subtasks, where no agent was involved. 
As shown in \cref{fig:defect-catch}, most defects asserted in the text of the artifact were raised by more than half of the participants; for example, the error to filter out non-adopters of Copilot from the analysis was caught by 70\% of participants. 
On the contrary, defects that require checking a claim against its evidence were raised by almost nobody. For example, "null result never stated" reported near-zero correlations of -0.005, but the interpretation text instead describes what a negative correlation would mean. None of the participants caught this error buried in the output that requires checking claims against the evidence.

\subsubsection{If the target is understanding: work incrementally and ask questions}
\label{sec:rq3-understanding}

All seven understanding outcomes come from the question-asking (both during planning and review) or critical-guidance paths (\cref{tab:strategy-outcome}). 
Two participants describe the technique to build understanding incrementally. P10 framed incremental complexity as a human requirement rather than a model one: \quoteinline{it's more kind of for the human interpretability to start from light and then incrementally [increase complexity]\ldots now you can give the agent more freedom, and it can do it as complex as needed}. P9 described decomposition for the same purpose --- \quoteinline{the agent performs better if you give it the task step by step instead of giving everything at once. And then if I do it step by step, then I will be able to understand what it's doing} --- with the failure mode it avoids stated directly: \quoteinline{if it generates a lot of code at once, then it becomes difficult for me to go through it and understand}.
P20 also adopted a simplicity principle to produce that understanding: \quoteinline{I'm gonna apply the Occam's razor principle here and just go with the most straightforward signal\ldots this strategy has worked best for me to also keep track of what's going on, consistently}.

Moreover, sessions ending in human understanding ($n=7$) average $23\%$ completeness and $62\%$ validity, and also participants landing in the understanding are mostly experts (5/7, 71\%).
Strategies for human understanding could cost completeness (\cref{sec:rq1-diligence}), and under a tight timeline, it can be very hard for novices to develop appropriate understanding.
For example, P17 (a novice designer) spent the session asking substantive conceptual questions such as what right-censoring means for an attrition label, but could not recall the specifics afterwards. 
Our study measures understanding as short-term recall, but long-term memory and \emph{learning} can take longer time or repetitions to develop (\cref{subsec:accelerate-explore}).
Question-asking may buy comprehension of the artifact at hand; but it does not, in one session, buy the domain knowledge that would let a novice build the next one.

\section{Discussion and Future Work}
\label{sec:discuss}

Our observations connect to how people work with code-generating agents, to how we should assess that work, and to how tools might adapt to the person behind the prompt.

\subsection{Slop and Sink vs. Stop and Think: {Fast and Slow Thinking in} AI-assisted Work}
\label{subsec:accelerate-explore}

Our results suggest one pathway through which AI-assisted work can become workslop \cite{liebscher2026workslop}: delegation can make the artifact advance faster than the worker's ability to understand and verify it. Task-relative expertise changes this risk, but does not eliminate it. Novices were especially vulnerable because they more often adopted completion-oriented workflows, delegating substantive decisions and reviewing the resulting artifact at the surface level. Yet experts could also end the task with little understanding when they delegated heavily or postponed inspection and underestimated agent-running time (\cref{sec:rq1-diligence}).

This perspective extends prior work's distinction between acceleration and exploration on AI-assisted programming \cite{Barke2022-tv} and resonates with the fast and slow thinking literature \cite{Kahneman2011-qi}. The fast workflow accelerates throughput, while the slow workflow, which includes decomposing the problem, questioning assumptions, and steering the agent, helps explore and build understanding (\cref{sec:rq3-understanding}).
Our study suggests that problems arise when acceleration becomes the default even when the worker lacks the task-relative expertise needed to slow down and supervise the result. 

Our data-analysis task, being open-ended and under-specified, enabled both fast and slow thinking, but not everyone could take up the \emph{slow strategy}. Higher-expertise participants could interrogate the agent's assumptions, probe the missing-label ambiguity, and use the interaction to refine \emph{what} the analysis should even be. Lower-expertise participants (even with low trust in AI) could be confined to acceleration with agents: they could drive the agent toward a complete-looking artifact, but lacked the domain knowledge to explore or verify.
We expected high-trust users to delegate to AI and accelerate, skipping exploration. But participants who reported high trust could revert to iterative exploration (especially experts) when the task was unfamiliar. 
Participants strongly believe understanding matters for downstream tasks, and even high-trust participants are unwilling to trade understanding for task completion under time constraints. 

As the agent rapidly expands the artifact, the amount of work the human must later understand and check also accumulates. Under bounded time, users may never catch up, leaving them with output that is easy to hand off but difficult to own. In this sense, an unpaid ``diligence tax" might turn into a ``verification debt" in the long run \cite{Xu2025-xh}.
Whether ``slow is fast'' in the long run, i.e., whether the understanding built now pays off in speed and quality later, is a question left for future work (\cref{subsec:limitations}), and it connects to a longer line of work on the trade-off between speed, quality, and learning in AI-assisted work \cite{He2026-tp}.

\paragraphBold{Implications for end-users: building expertise needs a necessarily slower workflow}

The practical advice for a worker whose goal is to grow, or who will have to defend the result, is to counteract the desire to automate: deliberately \emph{slow down}, iterate, ask, and verify rather than accept a finished file. 
Genuine upskilling requires sustained, active engagement, such as building with the method repeatedly, or focused training. Incidental explanation that accompanies a one-off task completion may just result in superficial parsing without understanding. 
For novices in unfamiliar task domains, developing understanding is harder and slower. Within our 30-minute task, even novices who iteratively engaged in question-asking with the AI could fail to develop durable understanding and were unable to recall the method post-task. Humans need repeated exposure to learn new concepts (e.g., seven practice opportunities to master a knowledge component \cite{Koedinger2023-xe}), and according to our participants, learning new skills often takes days or weeks (e.g., learning GitHub commands as a designer).

\subsection{Look Beyond Artifacts: Building Human Profiles from Interaction}
\label{subsec:inference-learning}

In our study (\cref{sec:rq2}), we observe how expertise and trust shaped \emph{how} people prompted, including which task dimensions they steered, where they questioned assumptions and made critiques, and what they left to defaults; interaction traces carry a readable signal of the worker's profile. 
For example, in our study's human-only subtasks, designers asked why, challenged the problem space, and focused on visuals; PMs asked about the deliverable and its audience; data scientists and software engineers went straight for the method. A profile inferred from behavior could therefore be quite fine-grained, distinguishing not just expert from novice but the \emph{axis} of a person's expertise.

Current models' and tools' default assumption is that the human is a capable supervisor or delegator who wants the whole task completed with minimal interruption \cite{Jagadeesh2026-tu}. However, this assumption can fail to support novices when they need more scaffolding, or even for experts to properly supervise agent work. In our study, out of all 15 participants with a model implemented (9 novices and 6 experts), 78\% of novices and 67\% of experts fail to realize the silent label circularity error buried in a complete-looking notebook (\cref{sec:rq3-validity}). 
A system optimized to hand over the whole output as fast as possible is, targeting task completion rather than human understanding, can be actively counterproductive: it maximizes the surface area for a silent error the user is least equipped to catch.

\paragraphBold{Implications for tool builders: infer expertise and reliance, and design for human oversight}
0As a worker's expertise is legible in how they interact (\cref{sec:rq2}), an agent can in principle sense, on the fly, whether it is working with an expert or a novice, and it should not respond agnostically. 
This opens a design space for \emph{profile-aware} support: tools that infer a user's expertise and disposition from their behavior and tailor scaffolding accordingly, for example, by surfacing the judgment calls a novice is skipping, suggesting the next question an expert would ask, or recommending tasks and prompts matched to the outcome the person is trying to reach. 
Such adaptation could help calibrate delegation, nudging over-delegating novices toward verification.
To design for human understanding and supervision in the loop, less is more: people need smaller increments, mandatory check-ins, and automatically triggered self-verification loops (or a foregrounded ``what to check'' surface) that do not assume the human will audit what comes back. 

Additionally, not all human characteristics are equally inferable from interaction traces. Expertise was relatively legible in our study through behaviors such as decomposition, methodological steering, and critical review (\cref{sec:rq2-workflow}). Trust, as an attitude, actually diverges from delegation reliance, which is the observed behavior (\cref{sec:rq2-saydo}). For example, a trusting expert may choose to delegate less and manually steer a lot when facing an unfamiliar task; a skeptical novice may have to delegate more than they prefer because they do not have the domain knowledge to provide any useful oversight.

For adaptive systems, this suggests that a system may be better positioned to infer whether a user is currently behaving in a hands-on or hands-off way, and whether their behavior exhibits task-relevant expertise, than to classify them as inherently “trusting” or “skeptical.” A hands-off expert and a hands-off novice may need different support, just as two equally trusting users may behave very differently depending on their task context, task familiarity, and expertise. Profile-aware agents should therefore condition support on observable interaction behavior and task-relative expertise.

\paragraphBold{Implication for assessment makers: look at the interaction and human, not just the artifact}
\label{subsec:assessment}

Our results also have direct implications for anyone using AI-assisted work products to assess people, instructors making take-home assignments, employers designing interview tasks, etc. 
As agent capacities keep improving, AI will increasingly close the gap in surface completion; the discriminating signal moves to what the worker understands and more nuanced quality: whether humans can justify their choices, answer follow-up questions, recognize what is missing or wrong, and extend the work. 
Assessment should therefore collect beyond the artifact to probe expertise and understanding directly, e.g., from interaction logs and a short defense or a live discussion.

\subsection{Limitations and Future Work}
\label{subsec:limitations}

These are hypothesis-generating findings from a small ($N=20$), single-organization sample performing one data-analysis task with one agent under a time limit, and our expertise and trust measures are self-reported (\cref{subsec:profiles}). 
The patterns we describe should be read as directional and grounded in qualitative evidence rather than as confirmed effects. 
As the 30-minute main task left little room for extended plan iteration or output critique, the human-only subtasks cannot be validated as proxies for main-task interaction or outcomes. 

A natural next step is a follow-up in which participants continue the task offline and prepare to defend it in a stakeholder discussion, logging their interactions again, so that exploration, verification, and understanding can be observed over a longer horizon. 
Additionally, there can be different types of understanding developed during different stages of the workflow, such as understanding of the problem space, chosen methods, or calculated results. Future work could dive deeper into how different understandings trade off with task outcomes and are shaped by the interactions. 
Future work should also observe iteration and evaluation behaviors that a short lab session cannot elicit and to test whether isolated-subtask skill predicts full-task performance at larger scale.

Establishing which behaviors drive outcomes, and whether profile-aware scaffolding actually helps people work or learn better, will require larger, longer, and more varied studies.
Our study provides a hypothesis-generating exploration of the impact of expertise and trust factors on AI-assisted work task and human outcomes. To better map out the whole landscape of AI-assisted work inputs and outputs (as portrayed in \cref{fig:framework}), future reviews or empirical studies could systematically vary agent and task characteristics and measure other task, human, and agent outcomes.

\section{Conclusion}
\label{sec:conclusion}

In this work, we set out to examine how a worker's expertise and their trust in AI shape agent-assisted work, and we instantiated a framework linking human profiles, interaction, and outcomes in a think-aloud study of 20 professionals on a realistic data-analysis task. We observed that expertise mattered less for a finished, complete-looking artifact than for whether the result was valid and whether the worker could understand and defend it. Experts and novices worked in legibly different ways. Experts decomposed plans, steered the agent, and interrogated its output, while novices pasted and accepted what came back. Those workflows bought different things, as the path to understanding is distinct from the path to coverage. 
Our study and analysis open up interesting questions to further explore human factors and outcome variables in AI-assisted work. We believe that future tools and assessments should be built to better serve diverse human expertise and targeted needs.

\begin{acks}
Thanks to all the participants for the pilot and user study in this work. This work is funded by Microsoft. 
Thanks to Martin Chan for assisting with task and rubric creation. 
Qianou Ma is partially funded by research grants from Gates Foundation and the CMU Learning Engineering Institute.

\textbf{Acknowledgment of LLM Use.} We acknowledge the use of LLMs to assist with language editing during manuscript preparation and data analysis. Implementation details, including prompts and settings, are attached in the Appendix.
\end{acks}

\bibliographystyle{ACM-Reference-Format}
\bibliography{paperpile, ref}

\clearpage\newpage


\appendix
\onecolumn

\section{Task Materials}
\label{app:tasks}

\subsection{Main Task: Employee Attrition Model}
\label{app:main-task}
\begin{taskbox}

\textbf{Think-Aloud Instructions}

This is a research study task. You are not being evaluated---there are no right or wrong answers. We are interested in how you approach and work through the task using AI assistance. Please think out loud as you work. Narrate your reasoning, decisions, and any confusion. Your think-aloud is recorded and transcribed as part of the study.

\medskip

\textbf{Background}

This task is inspired by a real-world data science request. It is intentionally open-ended and underspecified in some areas---handling ambiguity is part of the task. You are expected to identify gaps, ambiguities, and make reasonable assumptions where needed.

\medskip

\textbf{Task}

You have been tasked with a PM request to design and create an MVP employee attrition model using Viva Insights dataset. The goal is to build a prototype that will inform whether this feature is worth investing engineering resources.

The PM is interested in understanding:
\begin{itemize}
    \item If we were to create an employee attrition model, what would this look like?
    \item What would be the inputs and outputs of the model?
    \item What factors should we take into account when thinking about bringing this model into production?
\end{itemize}

\textbf{Data}

A starter sample Person Query dataset (\texttt{pq\_data}) is available via the \task{vivainsights} Python package (installed via: \texttt{pip install vivainsights}) if you haven’t.

\emph{Note:} This dataset is known to be incomplete. An attrition label column is missing and will need to be engineered or simulated---identifying and addressing this gap is part of the task.

\medskip

\textbf{Required Output}

Produce a Jupyter Notebook \texttt{main\_task.ipynb} that includes:
\begin{itemize}
    \item Your recommended choice of attrition model algorithm and rationale
    \item Code demonstrating how the model would run end-to-end
    \item Example outputs from the model (predictions, feature importance, evaluation metrics)
    \item A brief discussion of production considerations
\end{itemize}

\textbf{Using the AI Agent}

You are expected to use GitHub Copilot agent (model: \texttt{claude-sonnet-4-6}) for this task.

Use the AI agent as you naturally would: ask questions, plan, generate code, debug, interpret results, write narrative, or anything else that helps you complete the task.

\end{taskbox}

\subsection{Subtask Context: Feature Adoption Analysis}
\label{app:subtask-context}

Both subtasks concern a second, distinct analysis: a PM request to analyze Copilot feature adoption
and its relationship to collaboration patterns in Viva Insights data. The PM wants to understand
which employee segments are adopting Copilot features (and which are not); whether there is a
relationship between Copilot usage and collaboration metrics (e.g., meeting hours, email hours); and
what a production-ready adoption dashboard or report would look like. The subtasks use the same
dataset as the main task; the columns described to participants as relevant are listed in
\cref{tab:columns}.

\begin{table}[hpbt]
\centering
\caption{Columns described to participants as relevant for the subtask analysis context.}
\label{tab:columns}
\begin{tabular}{@{}lll@{}}
\toprule
Column & Category & Description \\
\midrule
\texttt{Organization} & Demographic & Business unit / department (e.g., IT, Legal, HR, Finance) \\
\texttt{LevelDesignation} & Demographic & Seniority label (e.g., Junior IC, Senior IC, Senior Manager) \\
\texttt{Total\_Copilot\_active\_days} & Copilot usage & Days per week the employee actively used Copilot (1--7) \\
\texttt{Copilot\_actions\_taken\_in\_Teams} & Copilot usage & Count of Copilot actions taken inside Microsoft Teams \\
\texttt{Copilot\_actions\_taken\_in\_Outlook} & Copilot usage & Count of Copilot actions taken inside Outlook \\
\texttt{Collaboration\_hours} & Collaboration & Total weekly hours in any collaboration activity \\
\texttt{Meeting\_and\_call\_hours} & Collaboration & Weekly hours in meetings and calls \\
\texttt{Email\_hours} & Collaboration & Weekly hours reading and writing email \\
\bottomrule
\end{tabular}
\end{table}

\subsection{Subtask 1: Edit a Plan}
\label{app:subtask1}

\begin{taskbox}

\textbf{AI Use}: Do NOT use GitHub Copilot or any AI tool for this subtask. This is a human-only task.

\medskip

Your AI agent has generated an initial plan to complete the task. Is this a good plan?

\medskip

\textbf{Instructions}

\begin{enumerate}
    \item Open the downloaded file \texttt{subtask1\_plan.docx} from your task folder.

    \item Without using AI, edit the plan to improve it. Turn on \textbf{Track Changes}.

    \item You may, for example, fix incorrect reasoning, strengthen weak sections, add missing steps, remove irrelevant content, reorder sections, etc. You do not need to write any code for this task.

    \item When you are satisfied with your edits or time is called, save the file and stop.
\end{enumerate}

\end{taskbox}

\subsection{Subtask 2: Evaluate an Output}
\label{app:subtask2}

\begin{taskbox}

\textbf{AI Use}: Do NOT use GitHub Copilot or any AI tool for this subtask. This is a human-only task.

\medskip

Your AI agent has generated an initial output for the task. Is this a good output?

\medskip

\textbf{Instructions}

\begin{enumerate}
    \item Open the file \texttt{subtask2\_output.ipynb} from your task folder. Read through it---you do not need to run the code unless you want to.

    \item Without using AI, write your next prompt in \texttt{subtask2\_prompt.docx}, including your critique and suggestions or improvement plan for the next iteration.

    \item When you are satisfied with your critique and plan, or time is called, save the file and stop.
\end{enumerate}

\end{taskbox}

\section{Survey Instruments}
\label{app:surveys}

Several items are adapted from prior instruments measuring AI-use skill,
delegation, and trust~\cite{Shao2026-hp, Ma2026-cd}.

\paragraphBold{Shared matrix rows}
Several matrix questions are rated once for each of seven \emph{skill areas}, and one pre-session
matrix is rated for each of seven \emph{tools}:

\begin{itemize}[labelwidth=*,leftmargin=1.3em,align=left,topsep=0.2em,itemsep=0.1em]
  \item \textbf{Skill areas} (\cref{app:skills}):\label{app:skills}
    (1) Exploratory data analysis (EDA); (2) Statistical analysis \& inference; (3) Machine learning
    / model building for data science; (4) Data visualization; (5) Working with large / messy
    datasets; (6) Deriving business insights from data; (7) Communicating technical results to
    non-technical audiences \& data-driven storytelling.
  \item \textbf{Tools} (\cref{app:tools}):\label{app:tools}
    (1) Python; (2) Jupyter Notebook; (3) pandas; (4) matplotlib / seaborn / plotly;
    (5) scikit-learn; (6) Excel / Google Sheets; (7) GitHub Copilot or other AI coding assistant.
\end{itemize}

\paragraphBold{Response anchors}
\label{app:anchors}
The closed scales are:

\begin{itemize}[labelwidth=*,leftmargin=1.3em,align=left,topsep=0.2em,itemsep=0.1em]
  \item \emph{Years}: None; $<$1\,yr; 1--2\,yrs; 3--5\,yrs; 6--10\,yrs; $>$10\,yrs.
  \item \emph{Frequency}: Never; Rarely; Monthly; Weekly; Daily.
  \item \emph{Confidence}: Not confident at all; Slightly; Moderately; Very; Extremely confident.
  \item \emph{Agreement}: Strongly disagree; Somewhat disagree; Neutral; Somewhat agree; Strongly agree.
  \item \emph{Comfort/Trust}: Very uncomfortable / No trust at all; Somewhat uncomfortable / Slightly trusting; Neutral; Somewhat comfortable / Trusting; Very comfortable / Complete trust.
  \item \emph{Intensity}: Not at all; Slightly; Moderately; Very; Extremely.
\end{itemize}

\subsection{Human-Profile Factor Composition}
\label{app:profile-composition}

The human-profile factors (\cref{subsec:profiles}) are each computed as the mean of their
$z$-scored constituent items. Items are ordinally encoded (\cref{app:anchors}). 
These factors are internally reliable (\cref{tab:profiles}). Cronbach's $\alpha\ge0.80$ is considered reliable, while mean inter-item correlations $>0.2$ indicate items are representative of the same content domain; $>0.4$ means some items could be redundant \cite{Tavakol2011-dy,Piedmont2014-iq}. 
While in-domain expertise and delegation trust are moderately correlated ($r=0.48$)---skill and willingness-to-delegate are related but not interchangeable. 

The exact item membership is:

\begin{itemize}[labelwidth=*,leftmargin=1.3em,align=left,topsep=0.2em,itemsep=0.2em]
  \item \textbf{In-domain expertise} (28 items): for each of the seven skill areas
    (\cref{app:skills}), four items---\emph{years of experience} (\texttt{yrs}), \emph{frequency of
    performing the skill without AI} (\texttt{freqNoAI}), \emph{self-rated confidence in one's own
    ability} (\texttt{Hconf}), and \emph{frequency of using AI to perform the skill}
    (\texttt{freqAI})---giving $7\times4=28$ items. In pre-survey terms, these are the ``years of
    experience in each skill area,'' ``how frequently do you use each skill,'' ``how confident are
    you in your own ability,'' and ``how frequently do you use AI to perform each task'' matrices.
  \item \textbf{AI-delegation trust} (4 items): \emph{trust in AI for data-science tasks}
    (\texttt{att\_trust\_ai\_dstask}; ``I trust GenAI/LLMs to produce reliable outputs for data
    analysis tasks''); \emph{trust without oversight} (\texttt{att\_trust\_ai\_nooversight}; ``How
    much do you trust AI agents to execute tasks accurately without human oversight?'');
    \emph{comfort delegating data-science tasks} (\texttt{att\_delegate\_ds}; ``I am comfortable
    delegating data analysis tasks to an AI agent''); and \emph{comfort delegating high-stakes
    tasks} (\texttt{att\_delegate\_highstakes}; ``How comfortable are you delegating high-stakes
    professional tasks to an AI agent?'').
\end{itemize}

\subsection{Pre-Session Survey}
\label{app:pre-survey}

\begin{enumerate}[labelwidth=*,leftmargin=2.2em,align=left,topsep=0.2em,itemsep=0.2em]
  \item \textbf{Participant ID} (short text).
  \item \emph{Which of the following best describes your current job role?} --- Product Manager; Software Engineer; Designer; Data Scientist; Other.
  \item \emph{Years of experience in current role} --- 0--1\,yr; 1--3\,yrs; 3--5\,yrs; 5--10\,yrs; $>$10\,yrs.
  \item \emph{What is your age?} --- 18--24; 25--34; 35--44; 45--54; 55--64; 65+.
  \item \emph{Gender} --- Man; Non-binary/gender diverse; Woman; Self-Described; Prefer not to answer.
  \item \emph{Ethnicity} --- Asian; Black or African American; Hispanic or Latino; Native American or Alaska Native; Native Hawaiian or Other Pacific Islander; White; Other.
  \item \emph{How many years of experience do you have using each of the following tools?} --- \emph{Years} scale, for each tool (\cref{app:tools}).
  \item \emph{How many years of experience do you have in each of the following skill areas?} --- \emph{Years} scale, for each skill area (\cref{app:skills}).
  \item \emph{How frequently do you use each of the following skills in your work?} --- \emph{Frequency} scale, for each skill area (\cref{app:skills}).
  \item \emph{How confident are you in your own ability to perform each of the following tasks?} --- \emph{Confidence} scale, for each skill area (\cref{app:skills}).
  \item \emph{How frequently do you use AI to perform each of the following tasks?} --- \emph{Frequency} scale, for each skill area (\cref{app:skills}).
  \item \emph{How confident are you in an AI agent's ability to perform each of the following tasks?} --- \emph{Confidence} scale, for each skill area (\cref{app:skills}).
  \item \emph{How familiar are you with LLM products (e.g., ChatGPT, Claude, Google Gemini, etc.)?} --- I use them regularly; I have some experience using them; I have heard of them but don't know much about their functionalities; No, I've never heard of them.
  \item \emph{How frequently do you use the following AI tools in your work?} --- \emph{Frequency} scale, for: LLM coding assistants/agent (GitHub Copilot, Claude Code, etc.); General-purpose LLM chatbots/agent (Microsoft 365 Copilot, ChatGPT, Claude, Gemini, etc.).
  \item \emph{How much do you agree with each statement?} --- \emph{Agreement} scale, for:
    \begin{enumerate}[labelwidth=*,leftmargin=1.6em,align=left,topsep=0.1em,itemsep=0.1em,label=(\alph*)]
      \item I am confident in my ability to prompt GenAI/LLMs to do what I want.
      \item I rely on AI tools to improve the quality of my work.
      \item I can get LLMs to complete the data analysis tasks I instruct them to do.
      \item I am comfortable delegating data analysis tasks to an AI agent.
      \item Generally, I trust GenAI/LLMs to produce reliable outputs for data analysis tasks.
    \end{enumerate}
  \item \emph{How comfortable and trusting are you regarding AI agents handling professional tasks?} --- \emph{Comfort/Trust} scale, for:
    \begin{enumerate}[labelwidth=*,leftmargin=1.6em,align=left,topsep=0.1em,itemsep=0.1em,label=(\alph*)]
      \item How comfortable are you delegating high-stakes professional tasks to an AI agent?
      \item How much do you trust AI agents to execute tasks accurately without human oversight?
    \end{enumerate}
  \item \emph{Reflecting on your typical work tasks, which best describes the role AI agents can play in your job?}
    \begin{enumerate}[labelwidth=*,leftmargin=1.6em,align=left,topsep=0.1em,itemsep=0.1em,label=H\arabic*:]
      \item Autonomous --- AI agent handles the task entirely on its own.
      \item Minimal Input --- AI agent requires minimal human input for optimal performance.
      \item Partnership --- AI agent and human form an equal partnership, outperforming either alone.
      \item Human-Led --- AI agent requires human input to successfully complete the task.
      \item Full Involvement --- AI agent cannot function without continuous human involvement.
    \end{enumerate}
  \item \emph{Briefly describe how do you usually use AI agents in your work.} (long text)
  \item \emph{Anything else you would like the research team to know before starting?} (optional long text)
\end{enumerate}

\subsection{Post-Session Survey}
\label{app:post-survey}

\begin{enumerate}[labelwidth=*,leftmargin=2.2em,align=left,topsep=0.2em,itemsep=0.2em]
  \item \textbf{Participant ID} (short text).
  \item \emph{Please indicate the extent to which each question describes your experience:} --- \emph{Intensity} scale, for:
    \begin{enumerate}[labelwidth=*,leftmargin=1.6em,align=left,topsep=0.1em,itemsep=0.1em,label=(\alph*)]
      \item How mentally demanding or difficult were the tasks overall?
      \item How hard/effortful did you have to work to complete the tasks?
      \item How irritated, stressed, or frustrated did you feel during the tasks?
      \item How hurried or rushed was the pace of the tasks?
      \item How interesting and engaging did you find the task?
      \item How successful do you feel you were in completing the task?
      \item How confident are you that your final output is high quality?
      \item How satisfied are you with how you used AI during the task?
      \item How familiar are you with the task context (e.g., employee attrition model, Copilot usage analysis, vivainsights)?
      \item How much did you understand the final output notebook for the main task?
      \item How much did you understand the provided plan in subtask1 in the end?
      \item How much did you understand the provided output notebook in subtask2 in the end?
      \item How familiar are you with the type of task in the main task (perform data analysis using AI)?
      \item How familiar are you with the type of task in subtask1 (editing data analysis plan manually)?
      \item How familiar are you with the type of task in subtask2 (evaluating data analysis output manually)?
    \end{enumerate}
  \item \emph{How confident are you in your own ability to perform each of the following tasks?} --- \emph{Confidence} scale, for each skill area (\cref{app:skills}).
  \item \emph{How confident are you in an AI agent's ability to perform each of the following tasks?} --- \emph{Confidence} scale, for each skill area (\cref{app:skills}).
  \item \emph{How much do you agree with each statement?} --- \emph{Agreement} scale, for the same five statements as pre-session Q15.
  \item \emph{How comfortable and trusting are you regarding AI agents handling professional tasks?} --- \emph{Comfort/Trust} scale, for the same two items as pre-session Q16.
  \item \emph{Reflecting on your typical work tasks, which best describes the role AI agents can play in your job?} --- the same five options as pre-session Q17 (Autonomous; Minimal Input; Partnership; Human-Led; Full Involvement), shown without the H-prefixes.
  \item \emph{What was most challenging about working with AI on this task?} (long text)
  \item \emph{What would have helped you work more effectively with AI on this task?} (long text)
  \item \emph{What would you do differently if you did this type of task with AI again?} (long text)
  \item \emph{Any other comments or feedback for the research team?} (optional long text)
\end{enumerate}

\section{Annotation, Grading, and Verification Details}
\label{app:grading}

This appendix documents how the derived analysis variables were produced. The procedures differ in
kind: the expertise$\times$trust \emph{archetypes} are computed deterministically from the survey
factors (\cref{subsec:profiles}); the \emph{notebook outcome grades}
(completeness$\times$validity) are produced with an LLM and then verified by two authors; and the \emph{workflow codes} and \emph{observed-delegation levels} are coded directly by
two authors from the sessions (\cref{subsec:measures}).  \cref{app:irr} reports the inter-rater reliability, and
all three are human-verified and reconciled.

\paragraphBold{Model and parameters}
All LLM-based annotation and grading used a single Azure OpenAI deployment, queried through the
chat-completions API. \Cref{tab:gradingparams} lists the settings. We did not set a sampling
temperature (the deployment's default was used); rather than rely on sampling determinism, we
constrained the output to a JSON schema, anchored every rubric item to explicit scoring criteria,
and had human annotators verify the results (\cref{app:irr}).

\begin{table}[h]
\centering
\caption{LLM annotation/grading configuration.}
\label{tab:gradingparams}
\small
\begin{tabular}{@{}ll@{}}
\toprule
Setting & Value \\
\midrule
Provider / API        & Azure OpenAI, chat completions \\
Deployment (model)    & \texttt{gpt-5.6-terra} \\
API version           & \texttt{2024-12-01-preview} \\
Sampling temperature  & not set (deployment default) \\
Response format       & \texttt{json\_object} (schema-constrained) \\
Calls                 & one completion per notebook / per annotation unit; cached per participant \\
\bottomrule
\end{tabular}
\Description{Configuration of the LLM-assisted notebook-grading procedure. Grading uses an Azure OpenAI deployment of gpt-5.6-terra through the chat-completions API}
\end{table}

\paragraphBold{Workflow codes and observed-delegation levels}
The workflow characterization and the H1--H5 observed-delegation level are coded by two authors
directly from the think-aloud transcripts, screen recordings, and prompt trajectories, not by the
model; the procedure and its reliability are described in \cref{subsec:measures} and \cref{app:irr}.
The LLM deployment below is used only for the notebook outcome grades.

\paragraphBold{Main-task notebook grading}
Each notebook is graded in a single
chat completion per notebook. The grading instrument is the
completeness $\times$ validity rubric of \cref{tab:simplerubric} (eight items, each scored 0/1/2). For every item, the
model must return a structured score, a 1--2 sentence justification, and the notebook cell numbers
used as evidence. Two authors then split and verified every grade against the notebooks, each covering half the sample ($10$ notebooks, $80$ item scores), correcting the model where it made a mistake. {All completeness and validity results in the paper use these human-verified scores}; the model scores are retained only for the agreement/accuracy analysis in \cref{app:irr}.
The system prompt is reproduced below (the study brief and rubric JSON are interpolated where indicated).

\lstset{basicstyle=\footnotesize\ttfamily\raggedright,breaklines=true,breakatwhitespace=false,%
        columns=fullflexible,keepspaces=true,frame=none,xleftmargin=0pt,%
        aboveskip=3pt,belowskip=3pt,literate={’}{'}1 {‘}{`}1}

\begin{lstlisting}

You are an expert data-science instructor grading Jupyter notebooks from a
research study, strictly and consistently against a fixed rubric.

## Study context
<the study brief, interpolated here verbatim.>

## Rubric (8 items, each scored 0/1/2)
Two axes: 
"completeness" = artifact completeness (implementation coverage and executability of the brief's required outputs); 
"validity" = epistemic validity (how the missing-label ambiguity was handled and whether the evaluation and its framing are trustworthy). Polished-looking notebooks are not always well-reasoned - judge the two axes independently.

<the eight rubric items, interpolated here as a JSON array.>

## Grading rules
- Score every item exactly 0, 1, or 2:
  0 = none / missing / completely incorrect (the "0" anchor),
  1 = partial (the "1" anchor),
  2 = full credit (the "2" anchor clearly met).
- Items marked na_allowed may instead score "na", ONLY under that item's own stated condition - the conditions differ per item, so read each one. "na" is not a judgment of quality; it removes the item from that notebook's validity denominator. Prefer a 0 over "na" whenever the participant had the chance to address the item and did not: "na" means the item could not apply, not that the work was poor.
- Judge only what is present in the notebook transcript. No credit for intentions or for work done outside the notebook.
- Cell outputs are truncated and figures replaced by placeholders like
  "[image/png output: ...]"; treat placeholders as evidence the code ran, and judge the figure's purpose from the code.
- An empty or near-empty notebook scores "na" on every validity item whose condition it meets, and 0 on the completeness items.
- When torn between two anchors, assign the LOWER score unless the higher anchor is clearly and fully met.
- Where an item's description says some evidence belongs to a different item, do not double-credit it here.
- justification: 1-2 concrete sentences referencing cell numbers.
- evidence: the cell numbers (integers) supporting the score; [] if none.

## Output format
Respond with ONLY a JSON object, exactly this shape, one entry per rubric item (all 8 rubric_ids, same ids as above):
{"scores": [{"rubric_id": "<id>", "score": 0 | 1 | 2 | "na", "justification": "<1-2 sentences>", "evidence": [<cell numbers>]}, ...],
 "overall_comment": "<1-3 sentence holistic summary>"}
Grade the following notebook (participant <PID>).

===== BEGIN NOTEBOOK TRANSCRIPT =====
<NOTEBOOK TRANSCRIPT: cells numbered 1..N in order, each with its source; code cells also carry their outputs, truncated to 900 chars>
===== END NOTEBOOK TRANSCRIPT =====
\end{lstlisting}


\paragraphBold{Rubric: completeness $\times$ validity}
The eight items and their verbatim scoring anchors appear in \cref{tab:simplerubric}. Four completeness items mirror the brief's four required outputs; four validity items target the risks the task probes (the missing label, the label-construction
strategy, leakage/circularity, and evaluation methodology).

\begin{table}[htbp]
\centering
\caption{The primary completeness$\times$validity rubric: eight items, each scored 0 (none /
missing / completely incorrect), 1 (partial), or 2 (full credit). Each validity item may instead be
marked \emph{n/a} under the condition stated in its row, in which case it leaves that notebook's
validity denominator (e.g., modeling and evaluation validity when no model was ever trained).}
\label{tab:simplerubric}
\footnotesize
\setlength{\tabcolsep}{4pt}
\renewcommand{\arraystretch}{1.1}
\begin{tabular}{@{}p{0.1\linewidth}p{0.3\linewidth}p{0.59\linewidth}@{}}
\toprule
Item (axis) & Criterion & Scoring anchors \\
\midrule
\texttt{algo\_rationale} \newline (completeness) &
Recommended choice of attrition-model algorithm and rationale; the recommendation must be stated, not just implied by code. &
\textbf{0:} no algorithm recommendation at all. \newline
\textbf{1:} an algorithm is chosen/used, or models are compared in code, but the reasoning is unstated or a generic one-liner. \newline
\textbf{2:} clear stated recommendation with substantive rationale (interpretability, performance, simplicity, MVP suitability), ideally acknowledging alternatives. \\
\addlinespace
\texttt{e2e\_execution} \newline (completeness) &
Code demonstrating how the model would run end-to-end, executed: cells show outputs with no unresolved errors in the pipeline path (executability only; what is demonstrated is \texttt{example\_outputs}). &
\textbf{0:} no working pipeline, largely unexecuted, or execution stops early at an unresolved error. \newline
\textbf{1:} pipeline present but with gaps: a core stage (load/label/features/train/evaluate) missing or vestigial, stale/unexecuted cells, or unresolved errors in side cells. \newline
\textbf{2:} full pipeline (load $\to$ label $\to$ features $\to$ train $\to$ evaluate) executed with visible outputs and no unresolved errors. \\
\addlinespace
\texttt{example\_outputs} \newline (completeness) &
Example outputs from the model: employee-level predictions (probabilities/risk scores), feature importance, and evaluation metrics (what is demonstrated, not whether the code ran---that is \texttt{e2e\_execution}). &
\textbf{0:} no model outputs shown. \newline
\textbf{1:} one or two of the three output kinds shown. \newline
\textbf{2:} all three shown: employee-level predictions, feature importance, and evaluation metrics. \\
\addlinespace
\texttt{production\_} \texttt{discussion} \newline (completeness) &
Discussion of production considerations. &
\textbf{0:} no production discussion. \newline
\textbf{1:} brief or generic mentions touching only one or two areas. \newline
\textbf{2:} substantive discussion across several areas (obtaining real labels, privacy/ethics/fairness, monitoring/retraining, deployment, human oversight). \\
\midrule
\texttt{data\_validity} \newline (validity) &
Explicitly notes the dataset lacks a true attrition label and states the implication/limitation of the simulated/proxy label (nearly everyone mentions the gap; the discriminating part is stating the implication). &
\textbf{0:} never acknowledges the missing label. \newline
\textbf{1:} mentions the gap or a generic ``this is simulated'' caveat without stating what that invalidates. \newline
\textbf{2:} states the gap AND its limitation: results are prototype-only, metrics do not reflect real attrition signal, and real HRIS labels are needed before trusting outputs. \newline
\textbf{n/a:} only when the notebook contains no engagement with the data at all (empty or abandoned before any inspection). \\
\addlinespace
\texttt{label\_validity} \newline (validity) &
Constructs a justified proxy/engineered label: behaviorally plausible assumptions grounded in actual dataset columns (e.g., high after-hours load or low collaboration raising risk); randomness/noise rather than a purely deterministic rule; a plausible base rate for the stated horizon (roughly 5--20\%). &
\textbf{0:} no label constructed; purely random assignment with no behavioral grounding; or nonsensical/inverted assumptions with no justification. \newline
\textbf{1:} label built but purely deterministic (a model can re-derive it exactly), weakly justified assumptions, or a base rate outside the 5--20\% range without justification. \newline
\textbf{2:} plausible column-grounded assumptions, with noise where a deterministic rule would otherwise be exactly recoverable, and a base rate in roughly the 5--20\% range. \newline
\textbf{n/a:} only when the notebook never attempted to construct or simulate a label (deciding \emph{not} to build one and saying why is a scoreable choice, not n/a). \\
\addlinespace
\texttt{model\_validity} \newline (validity) &
Avoids or squarely addresses leakage and circularity: does not present a model trained on a label derived from its own features as having learned real attrition, and excludes or flags features that trivially encode the label. Dropping a composite risk score while keeping the features it was computed from does \emph{not} count; temporal separation (training on prior weeks to predict a held-out later week) is sound and is \emph{not} penalized. &
\textbf{0:} circular setup presented as a real result---the label is derived from the same-period features the model is given (including dropping the composite score but keeping its direct inputs). \newline
\textbf{1:} circularity acknowledged but near-perfect metrics still framed as meaningful, or only partial feature hygiene. \newline
\textbf{2:} explicitly managed: excludes/flags label-defining features, and/or separates the independent and dependent variables by time. \newline
\textbf{n/a:} only when the notebook never trained a model. \\
\addlinespace
\texttt{eval\_validity} \newline (validity) &
Methodologically sound evaluation: a held-out split or cross-validation, no train/test contamination (e.g., fitting scalers or selecting features on the full data), and metrics appropriate to the class balance rather than accuracy alone. &
\textbf{0:} no held-out evaluation, or the model is evaluated on data it trained on. \newline
\textbf{1:} a split exists but flawed: preprocessing fit before the split, no stratification with a rare label, or accuracy-only reporting on imbalanced data. \newline
\textbf{2:} clean held-out or cross-validated evaluation with metrics suited to the class balance. \newline
\textbf{n/a:} only when the notebook never trained a model. \\
\bottomrule
\end{tabular}
\Description{Eight-item rubric for grading notebook completeness and epistemic validity.}
\end{table}

\subsection{Inter-Rater Reliability and Human Verification}
\label{app:irr}

Three of the paper's analyses reported human verification: the completeness$\times$validity notebook grades, the holistic observed-delegation level, and the four-phase workflow codes. We report the
agreement below.

\paragraphBold{Completeness and validity (human-verified scores)}
Two authors split the 20 notebooks (10 each) and independently verified LLM scores all eight rubric items
against \cref{tab:simplerubric}, giving 160 item scores. Compared against the LLM's scores on the same 160 items, exact agreement is
$90.6\%$ (completeness $91.2\%$, validity $90.0\%$); the quadratic-weighted $\kappa$ is $0.88$, and the mean absolute difference is $0.12$. Disagreements are small and mildly optimistic on the model's side of the $144$ numeric (non-NA items): LLM scored higher than human on $10$, lower on $4$, and identical on the remaining $130$.

\paragraphBold{Observed delegation level (H1--H5)}
The holistic delegation level annotated on each session (the \emph{observed} axis of
\cref{fig:say-vs-do} and the correlations in \cref{sec:findings}) was coded independently by two authors for all 20 participants. On the ordinal H1--H5 scale, exact agreement was $50\%$ and
within-one-level agreement $100\%$ (quadratic-weighted $\kappa=0.65$)---disagreements were never more than one adjacent level. The two coders then adjudicated
every disagreement to a single reconciled level, and all reported observed-delegation results use the reconciled levels.

\paragraphBold{Workflow strategy codes}
The four-phase workflow codes behind the Sankey (\cref{fig:sankey-expertise}) and the strategy-outcome table (\cref{tab:strategy-outcome}) were coded independently by two authors for all 20 participants. Over the four phases (80 decisions), agreement was $91.2\%$ with Cohen's $\kappa=0.84$. By phase, agreement was $85\%$ 
($\kappa=0.77$) for plan, $90\%$ ($\kappa=0.78$) for implement, $100\%$ ($\kappa=1.0$) for review,
and $90\%$ ($\kappa=0.81$) for outcome. Disagreements were resolved to a single agreed code, which the analysis uses.

\subsection{Subtask Grading}
\label{app:subtask-rubric}

The analysis of the human-only subtasks uses the \emph{seeded defects}: weaknesses we planted in the
plan and the output notebook.
\Cref{tab:subtaskrubric} lists the eight seeded defects and, for each, the number of participants who raised it in each subtask and overall. 
Five defects appear in both the plan and the output notebook and three only in the notebook, so a defect not applicable to a subtask is marked ``--''.

Grading the plan subtask requires care about \emph{authorship} (the submission is the AI's plan
after the participant's tracked edits) and about \emph{evidence} (each catch must be tied to
words the participant actually wrote). An author reviewed the submitted artifacts (edited plan and critique) to annotate the seeded flaws flagged by each participant.

\begin{table}[t]
\centering
\caption{The eight seeded defects (the answer key for the human-only subtasks). \emph{Overall} is
the number of participants (of $N=20$) who raised the defect in \emph{either} subtask (the merged
catch rate plotted in \cref{fig:defect-catch}); \emph{s1}/\emph{s2} break it down by subtask~1 /
subtask~2. The first five defects are asserted in the artifact's text; the last three are detectable only by checking a claim against its evidence.} 
\label{tab:subtaskrubric}
\footnotesize
\setlength{\tabcolsep}{5pt}
\renewcommand{\arraystretch}{1.15}
\begin{tabular}{@{}p{0.74\linewidth}*{3}{>{\centering\arraybackslash}p{0.05\linewidth}}@{}}
\toprule
& \multicolumn{3}{c}{Caught by (of 20)} \\
\cmidrule(l){2-4}
Seeded defect & overall & s1 & s2 \\
\midrule
\textbf{Non-adopters filtered out of the data.} The rows with zero Copilot days are the comparison
group an adoption question needs; dropping them removes 519 of 10{,}500 rows and makes an adoption
rate uncomputable. \newline \texttt{m\_nonadopter\_exclusion} & 14 & 12 & 5 \\
\addlinespace
\textbf{One adoption metric only.} Total\_Copilot\_active\_days is treated as the whole of
``adoption,'' although Teams and Outlook action counts are loaded alongside it. \newline
\texttt{m\_single\_metric} & 12 & 9 & 8 \\
\addlinespace
\textbf{Segmented by organisation only.} Organization is the sole grouping variable, although
LevelDesignation is loaded and available. \newline \texttt{m\_org\_only\_segmentation} & 12 & 10 & 9 \\
\addlinespace
\textbf{$r>0.3$ treated as a conclusion.} An arbitrary threshold is written in as a decision rule,
and the notebook then offers a causal reading of a correlation. \newline
\texttt{m\_causal\_inference} & 5 & 4 & 2 \\
\addlinespace
\textbf{Bar chart of correlations as the stakeholder visual.} Three correlation coefficients
rendered as bars, named as the artifact a stakeholder will see. \newline \texttt{v\_chart\_type} &
8 & 4 & 4 \\
\midrule
\textbf{Auto-scaled axis inflates near-zero correlations.} All three correlations are within 0.005
of zero, but the bar chart sets no $y$ limits, so matplotlib scales the axis to roughly
$-0.009$--$0$ and the three bars appear dramatically different. The same happens in cell~7, where
organisation means that all sit near 2.2 are ranked as though the order carried information.
\newline \texttt{v\_axis\_scale} & 3 & -- & 3 \\
\addlinespace
\textbf{Code does not do what the surrounding text claims.} Two discrepancies in the Data
Preparation step: three columns are selected and then never used, and the markdown claims a cleaning
step the code does not perform. \newline \texttt{m\_impl\_bug} & 0 & -- & 0 \\
\addlinespace
\textbf{Near-zero result never stated in the write-up.} The correlations come back at about
$-0.005$, i.e.\ no relationship. The Interpretation section describes only what a high or a negative
correlation would have meant and never states what was actually found. \newline
\texttt{i\_null\_result} & 0 & -- & 0 \\
\bottomrule
\end{tabular}
\Description{Eight seeded defects used as the answer key for the human-only critique subtasks, together with the number of the 20 participants who detected each defect. The first five defects are visible from reading the artifact text and are detected by between 5 and 14 participants. The final three require checking claims against code, results, or visual evidence and are much harder to detect: three participants notice the misleading axis scale, while none detect the implementation-text contradiction or the unstated near-zero result.}
\end{table}

\end{document}